\documentclass[prl,twocolumn,amsmath,amssymb,superscriptaddress]{revtex4-2}
\renewcommand{\figurename}{\textbf{Fig.~\!}}
\usepackage[normalem]{ulem}
\usepackage{graphicx}% Include figure files
\usepackage{todonotes} % for comments
\usepackage{dcolumn}% Align table columns on decimal point
\usepackage{tabularx}

\usepackage{bm}% bold math
\usepackage{hyperref}% add hypertext capabilities
\usepackage{textcomp}
\hypersetup{
    colorlinks=true,
    linkcolor=[HTML]{0066CC},
    citecolor = [HTML]{0066CC},
    filecolor=magenta,      
    urlcolor=[HTML]{0066CC},
    pdftitle={Empowering high-dimensional quantum computing},
    pdfpagemode=FullScreen,
    }

\usepackage{xcolor}
\usepackage{multirow}
\usepackage{nicefrac}
\usepackage{qcircuit}
\definecolor{green}{HTML}{50C878}
\usepackage{physics}

\usepackage{siunitx}

\newcommand{\Hss}{\hat{H}_\mathrm{ss}} % KS
\newcommand{\Hd}{\hat{H}_\mathrm{d}} % KS
\newcommand{\Hc}{\hat{H}_\mathrm{c}} % KS
\newcommand{\Hq}{\hat{H}_\mathrm{q}} % KS
\newcommand{\Hfull}{\hat{H}_\mathrm{full}}
\newcommand{\Hrotc}{\hat{H}_\mathrm{rot,c}} % KS
\newcommand{\Hrotqd}{\hat{H}_\mathrm{rot,qd}} % KS
\newcommand{\Hrotfull}{\hat{H}_\mathrm{rot,full}} % KS
\DeclareMathOperator{\sgn}{sgn}

\begin{document}

% \preprint{APS/123-QED}

% \title{Extensible two-photon-induced interactions between qudits}% Force line breaks with \\
\title{Empowering high-dimensional quantum computing \\ by traversing the dual bosonic ladder}

\author{Long B. Nguyen}
% \email{longbnguyen@berkeley.edu}
\thanks{These authors contributed equally. \\ Correspondence to \url{longbnguyen@berkeley.edu} and \url{noahgoss@berkeley.edu}.}
\affiliation{Department of Physics, University of California, Berkeley, California 94720, USA}
\affiliation{Computational Research Division, Lawrence Berkeley National Laboratory, Berkeley, California 94720, USA}

\author{Noah Goss}
% \email{longbnguyen@berkeley.edu}
\thanks{These authors contributed equally. \\ Correspondence to \url{longbnguyen@berkeley.edu} and \url{noahgoss@berkeley.edu}.}
\affiliation{Department of Physics, University of California, Berkeley, California 94720, USA}
\affiliation{Computational Research Division, Lawrence Berkeley National Laboratory, Berkeley, California 94720, USA}

\author{Karthik Siva}
\thanks{Current address: IBM Quantum, IBM T.J. Watson Research Center, Yorktown Heights, NY 10598, USA}
\affiliation{Department of Physics, University of California, Berkeley, California 94720, USA}

\author{Yosep Kim}
\affiliation{Department of Physics, Korea University, Seoul 02841, Korea}

\author{\\Ed Younis}
\affiliation{Computational Research Division, Lawrence Berkeley National Laboratory, Berkeley, California 94720, USA}

\author{Bingcheng Qing}
\affiliation{Department of Physics, University of California, Berkeley, California 94720, USA}

\author{Akel Hashim}
\affiliation{Department of Physics, University of California, Berkeley, California 94720, USA}
\affiliation{Computational Research Division, Lawrence Berkeley National Laboratory, Berkeley, California 94720, USA}

% \author{Ravi K. Naik}
% \affiliation{Department of Physics, University of California, Berkeley, California 94720, USA}
% \affiliation{Computational Research Division, Lawrence Berkeley National Laboratory, Berkeley, California 94720, USA}

\author{David I. Santiago}
\affiliation{Department of Physics, University of California, Berkeley, California 94720, USA}
\affiliation{Computational Research Division, Lawrence Berkeley National Laboratory, Berkeley, California 94720, USA}

\author{Irfan Siddiqi}
% \email{iasiddiqi@lbl.gov}
\affiliation{Department of Physics, University of California, Berkeley, California 94720, USA}
\affiliation{Computational Research Division, Lawrence Berkeley National Laboratory, Berkeley, California 94720, USA}
% \affiliation{Materials Science Division, Lawrence Berkeley National Laboratory, Berkeley, California 94720, USA}


\begin{abstract}
    High-dimensional quantum information processing has emerged as a promising avenue to transcend hardware limitations and advance the frontiers of quantum technologies. Harnessing the untapped potential of the so-called qudits necessitates the development of quantum protocols beyond the established qubit methodologies. Here, we present a robust, hardware-efficient, and extensible approach for operating multidimensional solid-state systems using 
Raman-assisted two-photon interactions. To demonstrate its efficacy, we construct a set of multi-qubit operations, realize highly entangled multidimensional states including atomic squeezed states and Schrödinger cat states, and implement programmable entanglement distribution along a qudit array. Our work illuminates the quantum electrodynamics of strongly driven multi-qudit systems and provides the experimental foundation for the future development of high-dimensional quantum applications.

% Pushing the boundary of quantum technologies necessitates the consideration of high-dimensional platforms that extend beyond the established methodologies and insights associated with binary states. To harness the untapped potential of quantum computing and quantum simulation using these so-called qudits, it is essential to explore novel pathways for tailoring the interactions between them in a robust and programmable fashion. In this study, we present the successful realization of sideband interactions between qutrit levels through two-photon driving. Our research sheds light on potential undesired effects arising from intense driving and offers possible circumvention strategies. Additionally, we leverage this technique to construct high-fidelity multi-qubit gates which are otherwise challenging to implement, then explore approaches to adapt the protocol for applications in larger quantum devices. These findings offer valuable insights into the driven dynamics of multilevel systems and provide an exciting avenue for future innovations and discoveries in the realm of scalable quantum computing and quantum information processing using qudits.

% The implemented unitary can be used as the primitive to engineer expressive qudit gates, to enhance quantum circuit compilation, or to emulate energy-exchange processes in nature
\end{abstract}

\maketitle

%\tableofcontents

\vspace{1em}
\noindent\textbf{Introduction}

\noindent Superconducting circuits have gained recognition as a promising platform in the pursuit of quantum simulation and quantum information processing. Yet, scaling up existing architectures poses formidable challenges in terms of fabrication imperfection, constrained connectivity, and demanding instrumentation overhead. One key strategy to overcome the hardware limitation is to promote a qubit to a qudit and utilize higher energy levels already present in the system to encode or transmit quantum information~\cite{wang2020qudits}. In particular, it has been proposed that using qudits may simplify quantum circuit complexity~\cite{gokhale2019asymptotic, wang2020qudits}, allow the native quantum simulation of certain natural systems~\cite{gustafson2022noise}, increase the information capacity and noise resilience in quantum communication~\cite{cozzolino2019high}, and lead to more efficient logical qubit encoding~\cite{campbell2014enhanced,anwar2014fast,watson2015fast}.

In parallel with the developments of qudits using photons~\cite{kues2017chip,wang2018multidimensional}, cold atoms~\cite{davis2019photon}, and trapped ions~\cite{ringbauer2022universal}, recent works have highlighted the potential of superconducting circuits for high-dimensional quantum computing~\cite{blok2021quantum,goss2022high,luo2023experimental,liu2023performing,cao2023emulating}. However, the progress in superconducting qudits is constrained by a limited interaction toolkit, with the notable absence of scalable XY-type interactions. This impedes the utilization of superconducting qudits in advancing high-dimensional quantum computing.

% Such interactions are crucial in binary quantum systems, but their implementation in qudits faces inevitable obstacles such as spectral crowding and increased hardware complexity. 
In this work, we report an extensible approach to operate high-dimensional solid-state systems based on microwave-induced double-excitation hopping, and then showcase its utility by implementing high-fidelity multi-qubit gates, creating a suite of high-dimensional entangled states, and transferring entanglement across a qudit array. The protocol is platform-agnostic and is compatible with other systems. Notably, our results include the development of a theoretical framework that elucidates the two-photon dynamics in strongly driven coupled-qudit systems, a new method to construct multi-qubit gates entirely based on qutrit interactions, the realization of high-fidelity atomic squeezed states and Schrödinger cat states in qudits, and the synthesis of quantum circuits for high-dimensional operations.

\begin{figure*}[t]
    \includegraphics[width=\textwidth]{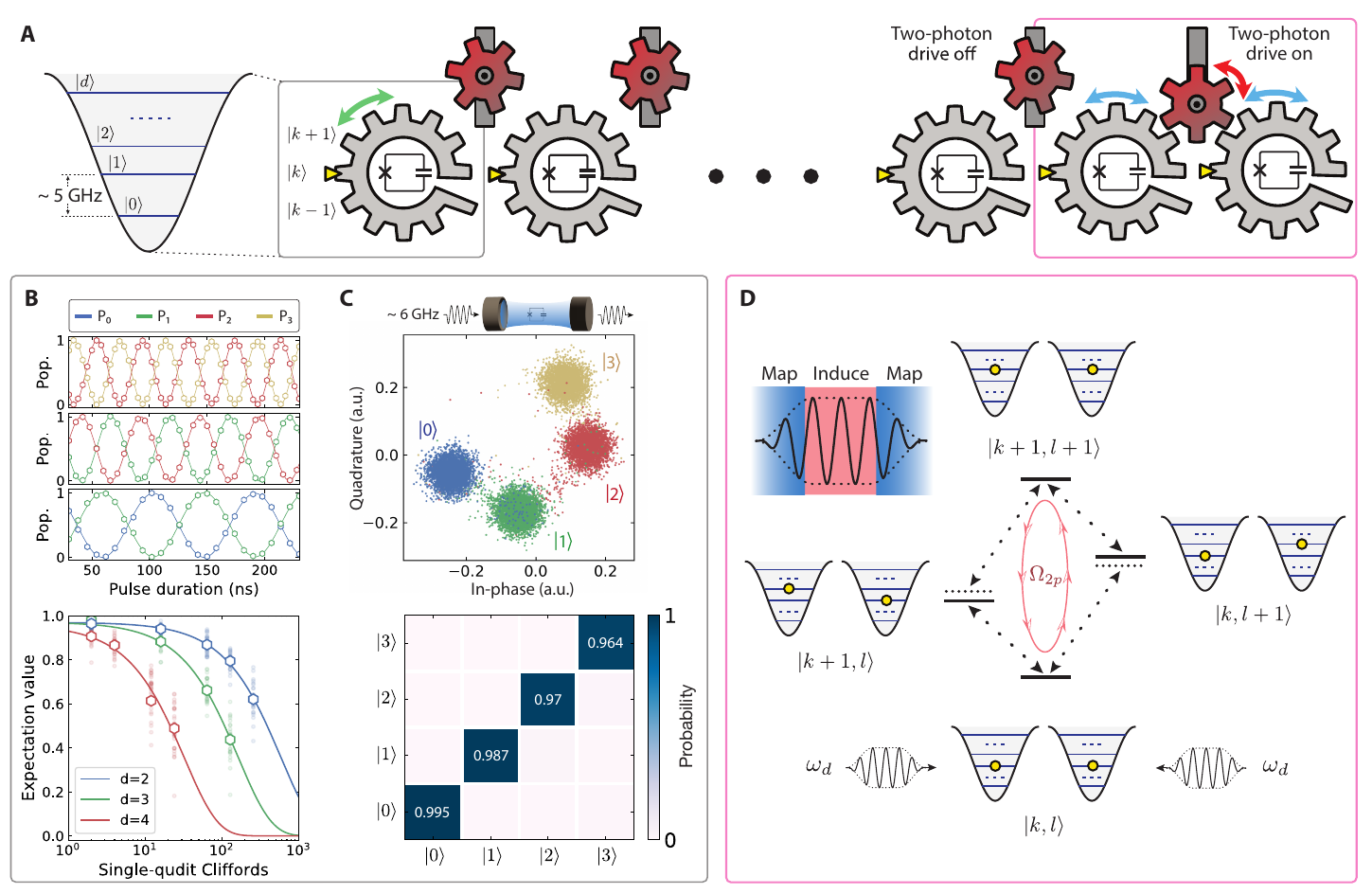}
    \caption{\label{fig1} \textbf{Overview of the experiment}. (\textbf{A}) Schematic of the high-dimensional system which is constructed by linking individual qudits into an array. Each qudit in the chain is a nonlinear harmonic oscillator comprising of a Josephson junction and a capacitor in parallel, and its eigenstates form a bosonic ladder of $d$ levels within the cosine potential. They are depicted as non-cyclical gears that can be rotated either individually using local controls or simultaneously using two-photon drives. (\textbf{B}) Local control of a qudit. (Top) Rabi oscillations of the populations under resonant microwave drives. (Bottom) Randomized benchmarking of single-qudit gates. (\textbf{C}) Readout of a qudit. (Top) A resonator is dispersively coupled to a qudit to measure its state. The probe signal from the resonator distinctly separates into individual blobs on the IQ plane, corresponding to the qudit being in the states $|0\rangle$, $|1\rangle$, $|2\rangle$, and $|3\rangle$. (Bottom) Preparation and measurement confusion matrix of qudit states, which reflects the readout fidelities. (\textbf{D}) Simplified depiction of the Raman-assisted two-photon-driven dynamics. The coupled-qudits pair form a set of eigenstates, visualized as a dual bosonic ladder. The entangling dynamic takes place in a four-level manifold within this structure. The effect of microwave drives applied to both qudits at a frequency close to the average of the single-excitation frequencies in this subspace is twofold: they first map the relevant qudit states to the dressed frame, then induce a Raman-assisted transition between the bipartite states $|k,l\rangle$ and  $|k+1,l+1\rangle$ (inset), resulting in an interaction at a rate $\Omega_{2p}$.}
\end{figure*}

\begin{figure*}[t]
    \includegraphics[width=0.98\textwidth]{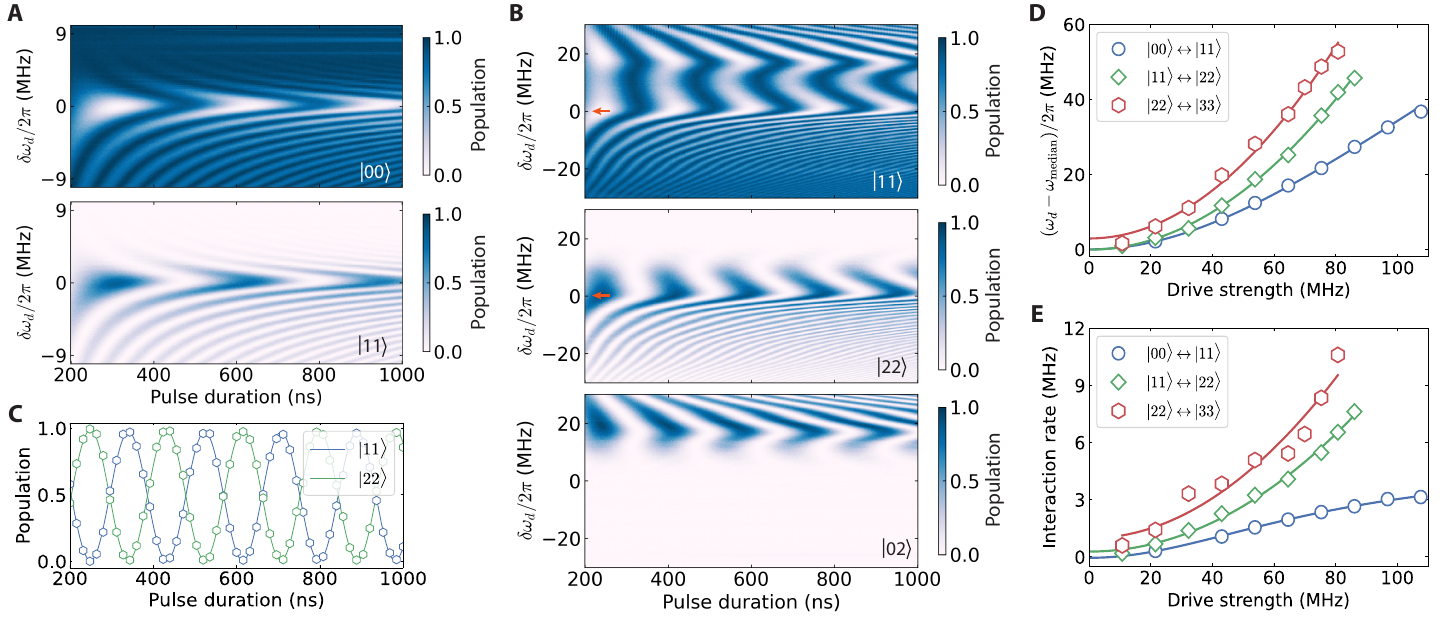}
    \caption{\label{fig2} \textbf{Traversing the dual bosonic ladder}. (\textbf{A}) Coherent $|00\rangle \leftrightarrow |11\rangle$ population exchange between two coupled qudits. The interaction is induced by monochromatic microwave drives, where the drive frequency is normalized with respect to the symmetric point. (\textbf{B}) Coherent $|11\rangle \leftrightarrow |22\rangle$ and $|11\rangle \leftrightarrow |02\rangle$ population exchange between two coupled qudits. The former interaction is induced by the two-photon Raman process, while the latter originates from the resonant condition between $|11\rangle$ and $|02\rangle$ in the driven frame. (\textbf{C}) Coherent flip-flop oscillation between $|11\rangle$ and $|22\rangle$ states obtained at the optimal drive frequency, indicated by the red arrows in panel \textbf{B}. (\textbf{D}) Optimal driving frequencies at various drive amplitudes. The y-axis is normalized with respect to $\omega_\mathrm{median}$, the median single-photon frequencies in the lab frame. The x-axis is expressed in terms of the single-photon Rabi rate in the 0-1 subspace. The solid lines are analytical fits. (\textbf{E}) Interaction rates of the dual bosonic transitions up to $d=4$ at various drive amplitudes. All data points are measured at the optimal driving frequencies. Analytical results are shown as solid lines.}
\end{figure*}

\begin{figure*}[]
    \includegraphics[width=0.95\textwidth]{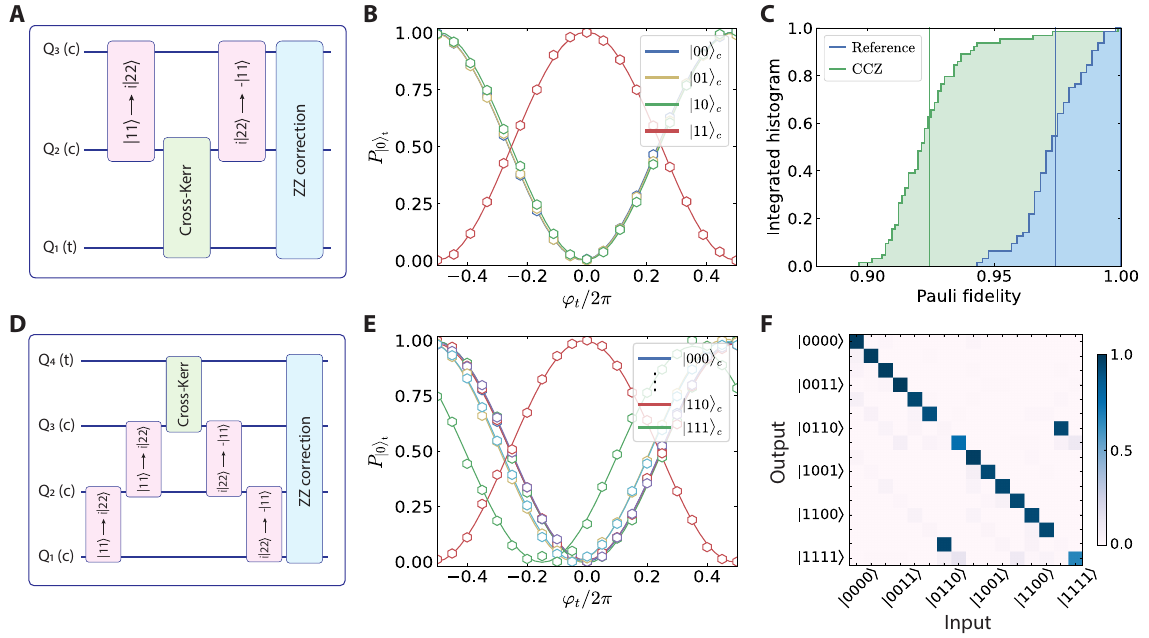}
\caption{\label{fig3} \textbf{Implementation and verification of multi-qubit gates}. 
(\textbf{A}) Gate sequence to implement a CCZ unitary. The qutrit swap gates (\textcolor{pink}{pink}) are used to shelve and then retrieve the control state $|11\rangle_c$. They sandwich a cross-Kerr gate (\textcolor{green}{green}) that induces a Z gate on Q$_1$ if and only if Q$_2$ is in $|2\rangle$. The final stage (\textcolor{cyan}{blue}) is used to correct the residual ZZ phases between the qubits. 
(\textbf{B}) The three-body operation manifests as the phase shift of the target qubit (Q$_1$) when the control qubits (Q$_2$ and Q$_3$) are in $|11\rangle_c$. (\textbf{C}) Pauli fidelities of the dressed cycle and the reference cycle from cycle benchmarking. The resulting gate fidelity is $\mathcal{F}_\mathrm{CCZ}=96.0(3)\%$. 
(\textbf{D}) Gate sequence to implement a CCCZ unitary. A cascade of qutrit swap interactions (\textcolor{pink}{pink}) is used to shelve and retrieve the respective $|11\rangle_c$ states. In the middle of the sequence is a cross-Kerr gate (\textcolor{green}{green}) that induces a Z gate on Q$_4$ if and only if Q$_3$ is in $|2\rangle$. All the residual correlated phases are corrected in the final stage (\textcolor{cyan}{blue}).\\
(\textbf{E}) The four-body operation is effectively revealed through a $\pi$-phase-shift of the target qubit (Q$_4$) for the control state $|110\rangle_c$.
(\textbf{F}) Truth table of the four-qubit Toffoli gate. The target state is shown to be flipped when the control state is $|$Q$_3$Q$_2$Q$_1\rangle=|110\rangle_c$ ($|0110\rangle \leftrightarrow |1110\rangle$). The corresponding truth table fidelity is $\mathcal{F}_\mathrm{CCCZ} = 92(1)\%$.}
\end{figure*}

\vspace{1em}

\noindent\textbf{Experimental Concepts}

\noindent A superconducting circuit constructed by shunting a Josephson junction with a large capacitor inherently behaves as a nonlinear harmonic oscillator with $d$ eigenlevels forming a bosonic ladder within the potential well~\cite{koch2007charge}. The circuit's eigenstates can be accessed using microwave drives to induce energy excitation, one level at a time. The qudits can thus be naturally visualized using noncyclical gears, as depicted in Fig.~\ref{fig1}\textbf{A}. A high-dimensional quantum device can be built by capacitively connecting the gears into an array, each one serving as a link in the chain with individual control and readout circuitry (Suppl. Mat. Note~1).

A snippet of the local control capability is shown in Fig.~\ref{fig1}\textbf{B}. On display are the Rabi oscillation data with near-perfect contrast and single-qudit randomized benchmarking results with average native gate fidelities of $\mathcal{F}_g=\{0.99936(3),0.99909(4),0.9978(2)\}$ for $d=\{2,3,4\}$, respectively. The observed Clifford fidelity decreases for higher $d$ due to the increasing number of native qudit rotations required, with compilations of Clifford groups requiring $\{2,6,12\}$ native gates for $d=\{2,3,4\}$, respectively. A dispersively coupled resonator allows the readout of qudit states via heterodyne measurement~\cite{PhysRevLett.105.223601}. As shown in Fig.~\ref{fig1}\textbf{C}, the readout signal separates into distinct blobs in the IQ plane for different qudit states, allowing high-fidelity initialization and measurement.

A natural approach to perform multi-qudit operations on these gears is to engineer their co-rotation which involves the simultaneous rotation of both qudits. This process embodies a non-energy-conserving two-photon dynamic characteristic of bosonic systems, depicted using the red idler gears in Fig.~\ref{fig1}\textbf{A}. We construct the desired interactions for all the high-dimensional subspace manifolds without additional circuitry or instrumentation as follows. 

First, a subspace spanned by the bipartite eigenstate $|k,l\rangle$ and its single- and double-excitation with negligible static cross-Kerr can be represented using the dual bosonic ladder as illustrated in Fig.~\ref{fig1}\textbf{D}. While transitions involving single-photon hopping are permitted, the direct two-photon hopping process $|k,l\rangle \rightarrow |k+1,l+1\rangle$ is prohibited due to parity. Notably, within the realm of quantum mechanics, a double excitation of this nature can take place through intermediate states. If these states are solely virtually occupied, the dynamics can be viewed as the multi-partite counterpart of the well-studied Raman process. In multi-qubit systems, this process has been utilized to engineer two-qubit gates in trapped ions with vibrational modes by employing two laser beams at frequencies that combine to match the double-excitation frequency~\cite{sorensen1999multiparticle}. Here, we harness these dynamics in solid-state systems without the need for ancillary components or additional degrees of freedom. 

To simplify the description, we examine the monochromatic driving scenario where two pulses are administered to both qudits at a single frequency $\omega_d$. This frequency is detuned by $\Delta_{1(2)} = \omega_d - \omega_{q,1(2)}$ from those of the relevant single-photon transitions $\omega_{q,1(2)}$, i.e.~between states $\ket{k,l}$ and $\ket{k+1,l}~(\ket{k,l+1})$ for qudit 1(2). In this context, the initial effect of the drives involves dressing the qudit states, contingent upon the fulfillment of the adiabaticity condition~\cite{nguyen2022programmable} (Fig.~\ref{fig1}\textbf{D}, inset). With a finite static coupling amplitude $g_{kl}$, the drives subsequently induce coherent two-photon hopping in the interaction frame via sequential single-photon virtual transitions through the hybridized states. Coherent transfer of the population between $\ket{k,l}$ and $\ket{k+1, l+1}$ ensues, with an oscillation rate $\Omega_\mathrm{2p}$ given as (see Suppl. Mat. Note~2)
\begin{equation}
\begin{split}
    \Omega_\mathrm{2p} = \frac{g_{kl}}{2}\bigg[&\left(1 + \cos \theta_1\right)\left(1 + \cos \theta_2\right) \\
    &+ \left(1 -\cos \theta_1\right)\left(1 - \cos \theta_2\right)\bigg],  
    \label{eq:swap_rate}
\end{split}
\end{equation}
where the angles $\{\theta_i\}$ are defined through the relation
   $ \cos \theta_i = \Delta_i/\sqrt{\Delta_i^2 + \Omega_i^2}.$
Here, $\Omega_{i}$ denotes the driving amplitude on qudit $i$, expressed in terms of the single-photon Rabi oscillation rate between the respective states.
% \begin{equation}
%     \Omega_{2p} = \frac{|\Omega_\mathrm{R_1}\Omega_\mathrm{R_2} - \Omega_{L_1}\Omega_\mathrm{L_2}|}{\Delta},
% \end{equation}
% where $\Omega_\mathrm{L,R}$ are the single-photon transition Rabi rates for on-resonant driving using the same pulses. 
% The interplay between these cascaded trajectories results in $\Omega_\mathrm{2p}=0$ for uncoupled qudits, where $\Omega_\mathrm{R(L)_1}=\Omega_\mathrm{L(R)_2}=\Omega_\mathrm{A(B),0}$. 
% and leads to a finite two-photon oscillation frequency given as (see Supplementary Materials)
% \begin{equation}
% \Omega_{2p}=2J\frac{\Omega_\mathrm{A,0}^2+\Omega_\mathrm{B,0}^2}{\Delta^2}
% \end{equation}

The monochromatic driving scheme is particularly well-suited for the current high-dimensional system, as superconducting qudits inherently possess finite detunings between their energy levels. The drive frequency can be adjusted by varying the ratio between the drive amplitudes, enabling the adjustment of the dressed qudit frequencies (Suppl.~Mat.~Note~2). Employing drives detuned similarly from the pertinent transitions also simplifies pulse calibration to ensure adiabaticity. 

It is crucial to highlight that the two-photon dynamics here resemble an ancilla-assisted tunneling process, in contrast with other high-order processes that rely on the system's nonlinearity~\cite{leghtas2015confining,grimm2020stabilization}. While the latter imposes stringent requirements on the system's frequencies and anharmonicity, the former is readily achievable given the existence of intermediate states that facilitate virtual transitions. The requirements for level detuning are relatively flexible, as a greater detuning permits stronger driving without inducing non-adiabatic effects. Therefore, the presented process is broadly feasible and adaptable across various platforms.

% The foundation of nonlinear and nonreciprocal devices is deeply rooted in high-order processes involving multiple quanta of energy. They have been utilized to realize transformative phenomena such as multi-wave-mixing~\cite{kamal2014asymmetric,naaman2022synthesis}, multi-photon emission and absorption~\cite{carusotto1968,zubairy1980}, up-conversion~\cite{deppe2008two}, and to engineer exotic quantum states such as cat states~\cite{leghtas2013hardware,grimm2020stabilization}. However, the strong drive amplitudes associated with these techniques have so far limited their applications in superconducting quantum qubits with passive encoding. A paradigmatic approach to leverage such a technique is to induce two-qubit interaction via two-photon drive~\cite{poletto2012entanglement,nesterov2021proposal}, which can be viewed as a sideband excitation~\cite{ann2022two} between two nonlinear systems. 

% We can view the dynamics via a simplified schematic as shown in Fig.~\ref{fig1}(a).

% So far, the theoretical framework above lacks a concrete experimental demonstration to verify its robustness and assess its potential for future quantum applications. To this end, we employ a transmon-based quantum device (Fig.~\ref{fig1}(b) and Appendix A), where all the components have fixed frequencies, making it the ideal testbed. The four qubits utilized in the experiments are labeled as Q$_{1-4}$, and their frequency allocation is shown in Fig.~\ref{fig1}(c). Each qubit is controlled and readout via designated coplanar transmission lines and resonators, respectively

\vspace{1em}

\noindent\textbf{Traversing the dual bosonic ladder}

\noindent We commence the experiment from the bottom of the ladder, starting with the $|00\rangle$ state. As the frequency $\omega_d$ of the microwave drives applied to the qudit pair is swept across the region around the median qudit frequency in the dressed frame, a coherent oscillation between $|00\rangle$ and $|11\rangle$ appears and forms a chevron pattern with respect to the drive frequency and pulse duration (Fig.~\ref{fig2}\textbf{A}). A pulse ramping time of $\tau_r=100~\mathrm{ns}$ is used to ensure adiabaticity~\cite{nguyen2022programmable}, which is verified by the absence of qudit population in the intermediate states following the drive.

The procedure is then extended to induce coherent energy exchange between other bipartite qudit states. Notably, within the $|11\rangle$ and $|22\rangle$
chevron, we observe an energy-conserving exchange interaction between $|11\rangle$ and $|02\rangle$ at certain drive frequencies, as shown in Fig.~\ref{fig2}\textbf{B}. This interaction is due to the resonance condition in the dressed frame between these states, previously reported in Ref.~\cite{nguyen2022programmable}, further validating our understanding of the strong drive effects in the system.  By detecting and avoiding such a condition, we can realize the desired interactions without unwanted effects. As Fig.~\ref{fig2}\textbf{C} shows, the fast oscillation between $|11\rangle$ and $|22\rangle$ at the optimal driving frequency (indicated by the red arrows in Fig.~\ref{fig2}\textbf{B}) remains highly coherent for longer than a microsecond without any spurious effect, indicating the robustness of the interaction. 

We then repeat the measurement at different drive amplitudes to thoroughly explore the two-photon drive dynamics. We note that microwave line crosstalk and the presence of other energy levels lead to a deviation of the optimal driving frequency from the median frequency between the pertinent multi-qudit levels, which can be accounted for by our analytical model (Fig.~\ref{fig2}\textbf{D}). 

The population oscillation observed at the optimal drive frequency is then used to extract the interaction rate, as shown in Fig.~\ref{fig2}\textbf{E}. Our analytical model again provides an excellent fit to the experimental data. We note that the higher we climb up the ladder, the faster oscillations we observe. We attribute this to the larger matrix elements of the exchange interaction at higher qudit levels. The accelerated interaction crucially allows for shorter gate times that compensate for the reduced coherence times of these transitions (Suppl. Mat. Note~1).

We additionally emphasize that while the current discussion focuses on the two-photon dynamics in the $\ket{k,l}$-spanned subspace for equal excitation quanta in each qudit, $k=l$, the process is also applicable for the general case, $k\neq l$, provided there is coupling between the intermediate states $\ket{k+1,l}$ and $\ket{k,l+1}$. For example, it can be leveraged to induce the energy exchange between $\ket{01}$ and $\ket{12}$ in two coupled qutrits (Suppl. Mat. Note~3). This feature further substantiates the extensibility of the protocol.

% Despite the possible spurious $|11\rangle \leftrightarrow |02\rangle$ exchange dynamics, the $|11\rangle \leftrightarrow |22\rangle$ oscillation along the optimal drive frequency remains coherent (Figure ~\ref{fig2}(c)). Interestingly, the oscillation amplitudes do not show any visible decay up to $1~\mathrm{\mu s}$, inspite of the relatively short coherence time of the $|2\rangle$ states of the qubits (Appendix~\ref{sec:app_device}), indicating the robustness of the induced qutrit-qutrit interaction. By extracting the oscillation rates at various drive frequencies in both subspaces, we map out the relation between driving strength and interactions rate, shown in Fig.~\ref{fig2}(d). We note the visibly stronger $|11\rangle \leftrightarrow |22\rangle$ rates compared to $|00\rangle \leftrightarrow |11\rangle$, which can be elucidated by taking into account the strong hybridization between the qutrit states.

\newpage
\noindent\textbf{Synthesis of multi-partite operations}
  
\noindent The higher levels of the qudits can be utilized for shelving, allowing access to otherwise impossible operations~\cite{fedorov2011implementation,chu2023scalable,nguyen2022programmable}. Here, we present a new method to synthesize high-fidelity multi-qubit gates using the two-photon interactions. The gate sequence, shown in Fig.~\ref{fig3}\textbf{A}, involves sandwiching a cross-Kerr two-qutrit gate~\cite{goss2022high} on Q$_1$-Q$_2$ between two $|11\rangle \leftrightarrow |22\rangle$ swap gates on Q$_2$-Q$_3$. 

First, the swap dynamics is used to substitute and then retrieve the control state $|11\rangle_c$. Then, the implemented cross-Kerr unitary between Q$_1$ and Q$_2$ applies a $\pi$-phase twist to Q$_1$'s $0-1$ subspace if and only if Q$_2$ is in its $|2\rangle$ state, effectively leading to a phase shift of the tripartite input state $|111\rangle$. In addition, the off-resonant drives introduce spurious ZZ coupling in the qubit subspace, which we take an additional step to negate using dynamical cross-Kerr pulses~\cite{nguyen2022programmable,goss2022high}.

% It is in essence similar to the sequence utilizing $|11\rangle \leftrightarrow |02\rangle$ energy-exchange interaction~\cite{fedorov2011implementation, reed2012realization,chu2023scalable,nguyen2022programmable}, with the notable exception of exchanging Q$_2$'s $|1\rangle$ state with $|2\rangle$ instead of $|0\rangle$. This switching thus necessitates a cross-Kerr interaction that induces an entangling phase in the qutrit subspace spanned by $|0\rangle$ and $|2\rangle$ instead of the regular qubit ZZ gate.

The entangling effect can be verified using a modified Ramsey sequence to probe the phase $\varphi_t$ of Q$_1$ for different Q$_2$ and Q$_3$ states~\cite{nguyen2022programmable}. As we apply the CCZ gate sequence, the target phase $\varphi_t$ corresponding to the control state $|11\rangle_c$ displays a $\pi$ shift relative to the others, signifying the three-body entanglement (Fig.~\ref{fig3}\textbf{B}). We then employ cycle benchmarking~\cite{erhard2019characterizing} (CB) to characterize the precision of the realized operation. By comparing the dressed cycle fidelity with the reference cycle fidelity, as shown in Fig.~\ref{fig3}\textbf{C}, we estimate a three-qubit gate fidelity of 96.0(3)\%.

% We further add a microwave-induced interaction at the end of the sequence to completely cancel residual correlated ZZ errors arises from static couplings and microwave crosstalks. The realized unitary then amounts to a control-control-Z, where the phase of the target qubit is flipped only when the control qubits are in their excited states, hence the designated control state is $|11\rangle$. This is in contrast with earlier works where the designated control state is $|10\rangle$~\cite{fedorov2011implementation,reed2012realization,chu2023scalable,nguyen2022programmable}. We proceed to benchmark the performance of the gate using cycle benchmarking~\cite{erhard2019characterizing}. Comparing the fidelity of the dressed cycle to that of the reference cycle (Fig.~\ref{fig3}(c)), we extract a gate fidelity of $96.0(3)\%$. Besides the decoherence errors during the gate sequence, we attribute $1-2\%$ error to leakage out of the qutrit subspace. We believe that it is rooted from the strong microwave crosstalk in the test device, through which the cross-Kerr gate can induce $|2\rangle \rightarrow |3\rangle$ transitions in neighboring qubits.

To further demonstrate the extensibility of this approach, we implement a four-qubit entangling gate by cascading the shelving steps as shown in Fig.~\ref{fig3}\textbf{D}. To showcase its versatility, we presently place the cross-Kerr gate between Q$_3$-Q$_4$ and perform the shelve-retrieve operations on the other pairs. Together with the ZZ correction at the end, the sequence effectively applies a $\pi$-phase-shift to the input state $|\mathrm{Q_4Q_3Q_2Q_1}\rangle=|1110\rangle$. We again verified this through a conditionality measurement, as shown in Fig.~\ref{fig3}\textbf{E}. We note that there exists a coherent error that manifests as a small phase shift in $\varphi_t$ when the control qubits are in $|111\rangle_c$ state. Our numerical simulation indicates that the infidelity from this error is insignificant compared to the obtained result (Suppl. Mat. Note~4).

Since finding the inverse unitary is challenging for a four-qubit circuit, CB cannot be used to determine the gate fidelity efficiently. Alternatively, we validate the operation by sandwiching it between two Hadamard gates on Q$_4$ and measuring the qubit states for different combinations of input states. The truth table summarizing the outcomes in Fig.~\ref{fig3}\textbf{F} shows the flipping of the target qubit (Q$_4$) when the control state is $|\mathrm{Q_3Q_2Q_1}\rangle=|110\rangle$. The corresponding truth table fidelity is $\mathcal{F}_\mathrm{CCCZ}=92(1)\%$, on-par with the best results previously reported in superconducting qubits~\cite{chu2023scalable} and trapped-ions~\cite{katz2023demonstration}.

\begin{figure}[t]
    \includegraphics[width=0.47\textwidth]{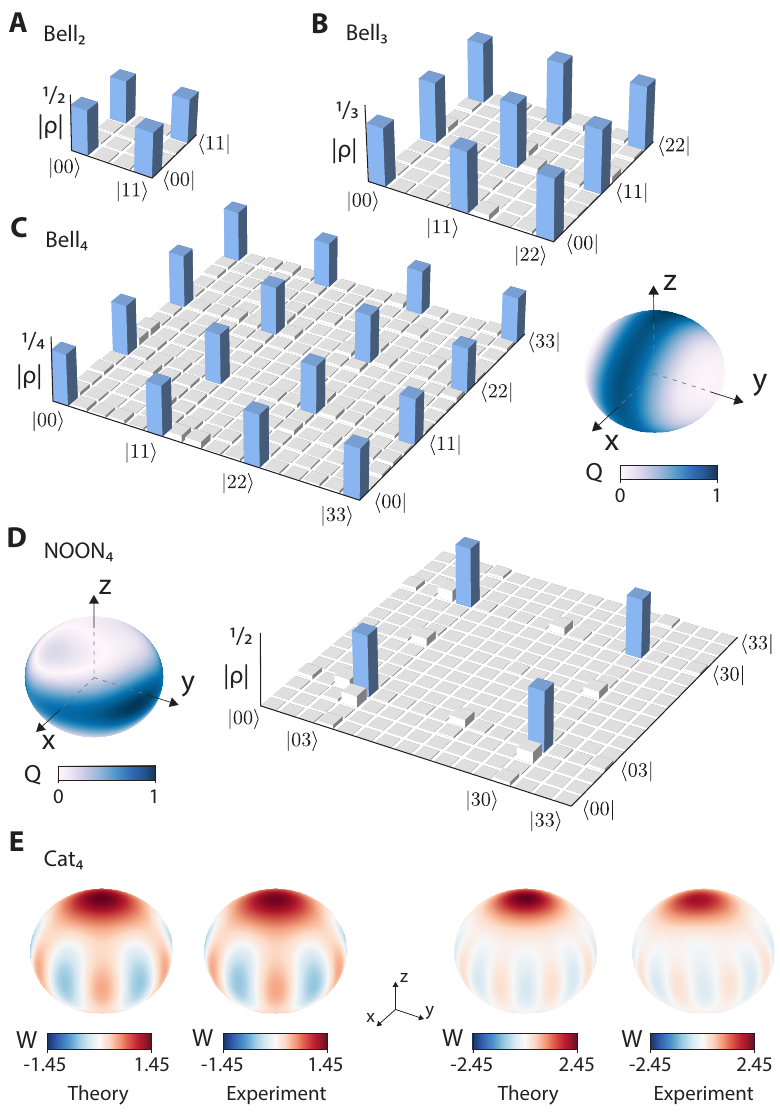}
\caption{\label{fig4} \textbf{High-dimensional entanglement}. (\textbf{A}) Density matrix of the qubit ($d=2$) Bell state with a raw (purified) fidelity of 99.2\% (99.9\%). (\textbf{B}) Density matrix of the qutrit ($d=3$) Bell state with $\mathcal{F}=97.7\%~(99.6\%)$.  (\textbf{C}) Density matrix (left) and Husimi-Q distribution (right) of the ququart ($d=4$) Bell state with $\mathcal{F}=94.3\%~(99.3\%)$. (\textbf{D}) Husimi-Q distribution  (left) and density matrix (right) of the ququart ($d=4$) NOON state $|30\rangle + e^{i\alpha}|03\rangle)/\sqrt{2}$ with $\mathcal{F}=94.6\%~(97.3\%)$. (\textbf{E}) Wigner functions of the high-dimensional atomic cat states, $(|00\rangle + e^{i\alpha}|33\rangle)/\sqrt{2}$ with $\mathcal{F}=98.6\%~(99.3\%)$ (left) and $(|000\rangle + e^{i\alpha}|333\rangle)/\sqrt{2}$ with $\mathcal{F}=80.1\%~ (90.9\%)$ (right).}
\end{figure}
\vspace{1em}

\noindent\textbf{High-dimensional quantum entanglement}

\noindent The capacity of a quantum system is often associated with how efficiently entanglement can be generated within it. The two-photon interaction is particularly well-suited to propagate entanglement to the higher levels in coupled qudit systems. By traversing the dual bosonic ladder, we proceed to create qudit Bell states with $d=\{2,3,4\}$. To verify our findings, we perform quantum state tomography on the output using SU(4) local rotations (Suppl. Mat. Note~5). The results are shown in Fig.~\ref{fig4}\textbf{A}-\textbf{C}. To estimate the effects of decoherence on the reconstructed states, we compute the fidelity with and without McWeeny purification \cite{mcweeny1960some}, which finds the nearest idempotent matrix. 
% Since the purified density matrix is a pure state, the remaining infidelity is due to either coherent gate errors or state-preparation and measurement (SPAM) errors
% We note that the errors caused by decoherence can be reduced by purifying the density matrices. 
This purification leads to a significant improvement of the quantum state fidelities compared to the unpurified results, highlighting that decoherence is a primary factor affecting the results. We observe a minor decrease in fidelity for purified states of larger dimension $d$, which is attributable to more state-preparation and measurement (SPAM) errors. We provide an estimation of tomographic uncertainty due to shot noise in Suppl. Mat. Note~6.

Notably, the cascaded two-photon interactions are reminiscent of the two-mode squeezing effect in quantum optics, which is described by the operator $\hat{\mathcal{S}}(\zeta)=\mathrm{exp}(\zeta\hat{a}^\dagger\hat{b}^\dagger - \zeta^*\hat{a}\hat{b})$, where $\zeta$ is the squeezing strength, and $\hat{a}$ ($\hat{b}$) is the annihilation operator on the first (second) mode. This results in a resemblance between the high-dimensional Bell states and photonic two-mode squeezed states, $|\Psi\rangle_\mathrm{2ms}\propto \sum_{N=0}^\infty c_N |NN\rangle$, where $c_N$ is a coefficient dependent on $\zeta$ and $N$ is the quanta number~\cite{caves1985new}. In the present system, $N$ effectively ranges from 0 to $d-1$.

We verify the squeezing effect by analyzing the Bell$_4$ state using the Husimi-Q quasiprobability distribution (QPD). Here, we use the collective spin coherent state as the basis state with isotropic QPD~\cite{arecchi1972atomic} (Suppl. Mat. Note~7). In contrast, a squeezed state would manifest as an elongated strip with a shrunk QPD along a geodesic on the sphere~\cite{kitagawa1993squeezed}. As shown in Fig.~\ref{fig4}\textbf{C}(right), the measured state indeed displays squeezing along the y-axis, confirming the expected outcome. Interestingly, the observed QPD resembles the results associated with atomic squeezed states generated by the two-axis countertwisting Hamiltonian~\cite{kitagawa1993squeezed}.

The NOON state is another important entangled state in quantum metrology, recognized for its exceptional sensitivity in measuring phase differences in interferometric experiments. It is defined as $|\Psi\rangle_\mathrm{NOON}=(|N0\rangle + e^{i\alpha}|0N\rangle)/\sqrt{2})$, where $N$ equals to $d-1$ in qudits. We proceed to utilize a combination of a two-photon process and single-qudit rotations to construct a high-fidelity NOON$_4$ state, as depicted in Fig.~\ref{fig4}\textbf{D}. Notably, the Husimi-Q QPD verifies that the NOON state exhibits squeezing along the geodesic orthogonal to that of the Bell states. 

Atomic Schrödinger cat states, which are superpositions of macroscopically distinct quantum states, belong to another category of entangled states that attract considerable interest~\cite{agarwal1997atomic}. In an $n$-qudit system where each qudit has dimension $d+1$, the coherent states of interest are $|00\dots 0\rangle$ and $|dd\dots d\rangle$~\cite{kitagawa1993squeezed}. Using the two-photon interaction, we examine two different high-dimensional atomic cat states, $|00\rangle + e^{i\alpha}|33\rangle$ and $|000\rangle + e^{i\alpha}|333\rangle$. To characterize these highly entangled states, we employ the Wigner function, which is known to allow the visualization of cat states due to its inherent capacity to accommodate negative values~\cite{rundle2017simple}. The experimental results in Fig.~\ref{fig4}\textbf{E} show the fringes characteristic of cat states, verifying our analytical prediction.\\

\begin{figure*}[]
    \includegraphics[width=0.8\textwidth]{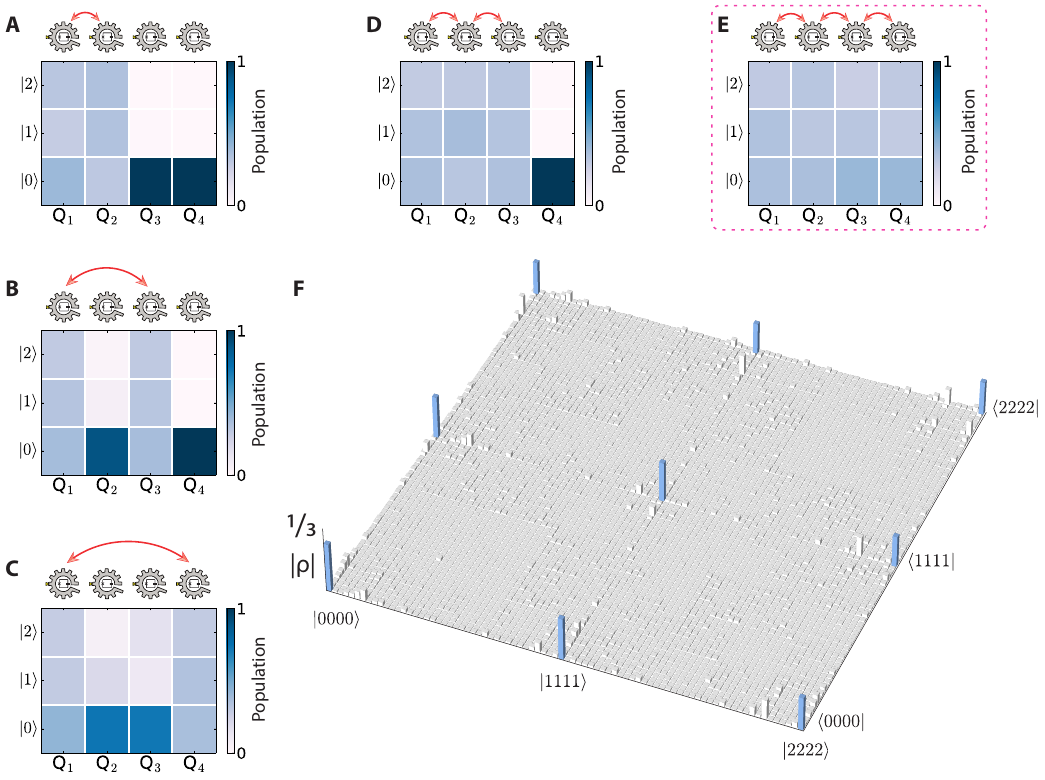}
\caption{\label{fig5} \textbf{Distribution of qudit entanglement}. Measured populations of (\textbf{A}) Q$_1$-Q$_2$ Bell$_3$ state, (\textbf{B}) Q$_1$-Q$_3$ Bell$_3$ state, (\textbf{C}) Q$_1$-Q$_4$ Bell$_3$ state, (\textbf{D}) Q$_1$-Q$_2$-Q$_3$ GHZ$_3$ state, (\textbf{E}) Q$_1$-Q$_2$-Q$_3$-Q$_4$ GHZ$_3$ state. (\textbf{F}) Density matrix of the four-qudit GHZ$_3$ state. The raw and purified fidelities are 70.6\% and 82.7\%, respectively.}
\end{figure*}

\noindent\textbf{Distribution of qudit entanglement}

\noindent The process of swapping populations between the qudit states $\ket{k,l}$ and $\ket{k+1,l+1}$ is functionally equivalent to a subspace energy-conserving swap process $\ket{k+1,l}\leftrightarrow\ket{k,l+1}$. This mechanism is crucial in constructing high-dimensional quantum circuits and is particularly important for distributing entanglement across a qudit array with limited connectivity. To examine this feature, our approach begins with the preparation of a two-qudit Bell$_3$ state (Fig.~\ref{fig5}\textbf{A}). Following this, we engage in a two-pronged approach: firstly, we focus on redistributing entanglement among distant qudits, and secondly, we aim to extend entanglement throughout the entire qudit array. This dual strategy explores the breadth of entanglement distribution possibilities in high-dimensional quantum systems with larger Hilbert space compared to architectures based on qubits.

To approximate the effectiveness of the operations, we directly measure the populations of the participating qudit states after implementing the sequences. This focuses on examining the diagonal elements of the multi-qudit density matrices and provides a straightforward approach to assess the impact of the swap operation. The observed transfer of population to adjacent qudits, as depicted in Fig.~\ref{fig5}\textbf{B,C}, confirms the successful redistribution of populations across the qudit array. 

Additionally, we verify their entanglement by measuring the fidelities of the resulting Bell$_3$ states between Q$_1$-Q$_3$ and Q$_1$-Q$_4$ (Suppl. Mat. Note~5). The results show quantum state fidelities of 75.7\% for the unpurified density matrices and 96.4\% for the purified ones in the case of Q$_1$-Q$_3$, and 53.0\% (unpurified) versus 95.7\% (purified) for Q$_1$-Q$_4$. The substantial discrepancy between the unpurified and purified results indicates that decoherence is the primary source of error in the operations.

Notably, we find that expanding the entanglement space to encompass other qudits requires fewer subspace swaps, in comparison with entanglement redistribution. This leads to the efficient generation of multi-qudit GHZ$_3$ states (Suppl. Mat. Note~8). The population measurement results show minimal imperfections, as illustrated in Fig.\ref{fig5}\textbf{D,E}. 

We proceed to measure the four-qudit GHZ$_3$ state achieving a fidelity of 70.6\% (82.7\%) for unpurified (purified) density matrix, which is shown in Fig.~\ref{fig5}\textbf{F}. While the difference between these metrics originates from unavoidable decoherence, we attribute the infidelity of the purified result to SPAM errors occurring during the long tomography sequences. In comparison, the three-qudit GHZ$_3$ unpurified (purified) density matrix has a fidelity of 90.8\% (96.4\%) (see Suppl. Mat. Note~5). This implies the need for more efficient and SPAM-free verification methods and benchmarking tools in the future, which we expect to provide a more accurate assessment of the robustness of our procedure.

\vspace{1em}

\noindent\textbf{Discussion}

\noindent We systematically showcase the operational principles of a multi-qudit device based on Raman-assisted two-photon interactions. In contrast with other dynamical processes in strongly driven systems, the reported quantum dynamics remain largely independent of nonlinearity~\cite{leghtas2015confining,grimm2020stabilization} or level-matching~\cite{poletto2012entanglement}. In addition, the optimal drive frequency can be tuned \textit{in situ} by adjusting the drive amplitudes and their relative ratio. These advantages enhance the extensibility and adaptability of the process, providing a crucial interaction repertoire to high-dimensional solid-state systems. Our findings lay the groundwork for future investigations into superconducting qudits, drawing insights from both bosonic and atomic research advancements.

% providing a significant boost to the high-dimensional interaction repertoire. We further demonstrate its power and versatility by implementing and characterizing a suite of high-fidelity multiqubit gates and high-dimensional entangled states with $d$ up to 4. 
% The implications of our work position superconducting circuits at the forefront of quantum computing with qudits.

As the simplicity and robustness of fixed-frequency superconducting circuits with fixed coupling have allowed them to be scaled up to devices integrating hundreds of qubits, our demonstrations using the same type of device without additional circuitry or instrumentation represent a potentially transformative approach to advancing quantum technologies through high-dimensional quantum operations. On the one hand, this work opens the door for exploring quantum sensing and quantum information processing with high-dimensional systems, and on the other hand, the results encourage the future development of bosonic encoding using superconducting qudits. Concurrently, the versatility of the protocol motivates its adaptation to implement highly connected quantum architectures. Looking forward, the observed decoherence-limited performance metrics underscore the necessity of developing hardware with enhanced noise protection for higher levels.\\

% perform high-dimensional quantum computing and quantum information processing tasks. We anticipate that our results will expedite the integration of high-dimensional frameworks into existing quantum systems, boosting research activity in this direction. Moreover, the achieved qudit entanglement holds promise for enhancing quantum sensing and quantum communication applications.

%%%%%%%%%%%%%%%%%%%%%%%%%%%%%%%%%%%%%%%%%%%%%%%%%%%%%
%               Acknowledgement
%%%%%%%%%%%%%%%%%%%%%%%%%%%%%%%%%%%%%%%%%%%%%%%%%%%%%
\noindent\textbf{Acknowledgments}

\noindent We thank Liang Jiang, Konstantin Nesterov, Seth Merkel, and Ken Xuan Wei for valuable discussions. This work was supported by the Office of Advanced Scientific Computing Research, Testbeds for Science program, Accelerated
Research in Quantum Computing Program, Office of Science of the U.S. Department of Energy under Contract No. DE-AC02-05CH11231. Yosep Kim acknowledges funding from the Creation of the Quantum Information Science R\&D Ecosystem through the National Research Foundation of Korea (NRF), grant No. 2022M3H3A106307411.\\

\noindent\textbf{Author contributions}

\noindent L.B.N.~conceptualized and organized the experiment. N.G.~established the high-dimensional metrology and calibration framework. L.B.N.~and N.G.~acquired the data and analyzed the results. K.S.~developed the theoretical model. Y.K.~assisted with visualization and devised the quasiprobability distribution analysis. E.Y.~developed the algorithms to optimize the high-dimensional circuits. B.C.~simulated the tomography shot noise, assisted with the Wigner function analysis, and provided the device image. A.H., D.I.S., and I.S.~oversaw the experimental effort. All authors contributed to the writing and editing of the manuscript. \\

\noindent\textbf{Competing interest}

\noindent The authors declare no competing interests.\\

\noindent\textbf{Data availability}

\noindent All data are available.
%%%%%%%%%%%%%%%%%%%%%%%%%%%%%%%%%%%%%%%%%%%%%%%%%%%%%
%               Appendices
%%%%%%%%%%%%%%%%%%%%%%%%%%%%%%%%%%%%%%%%%%%%%%%%%%%%%   
% \section*{Appendices}
% \input{sections/a1_Device}

\bibliography{apssamp}% Produces the bibliography via BibTeX.

\newpage
\clearpage

\renewcommand{\figurename}{Figure \!S\!\!}
\renewcommand{\theequation}{S\arabic{equation}}
\setcounter{page}{1}
\setcounter{figure}{0}
\setcounter{equation}{0}

\onecolumngrid
\vspace{\columnsep}
\begin{center}
\large \textbf{Supplementary Materials}
\end{center}
\vspace{\columnsep}
\twocolumngrid
% \begin{widetext}
% \centering
%     \large \textbf{Supplementary Materials \\ for XYZ}
% \end{widetext}

\subsection{Supplementary Note 1 --
Experimental Device}
% \subsubsection{Device description}
\begin{figure*}[t]
    \includegraphics[width=\textwidth]{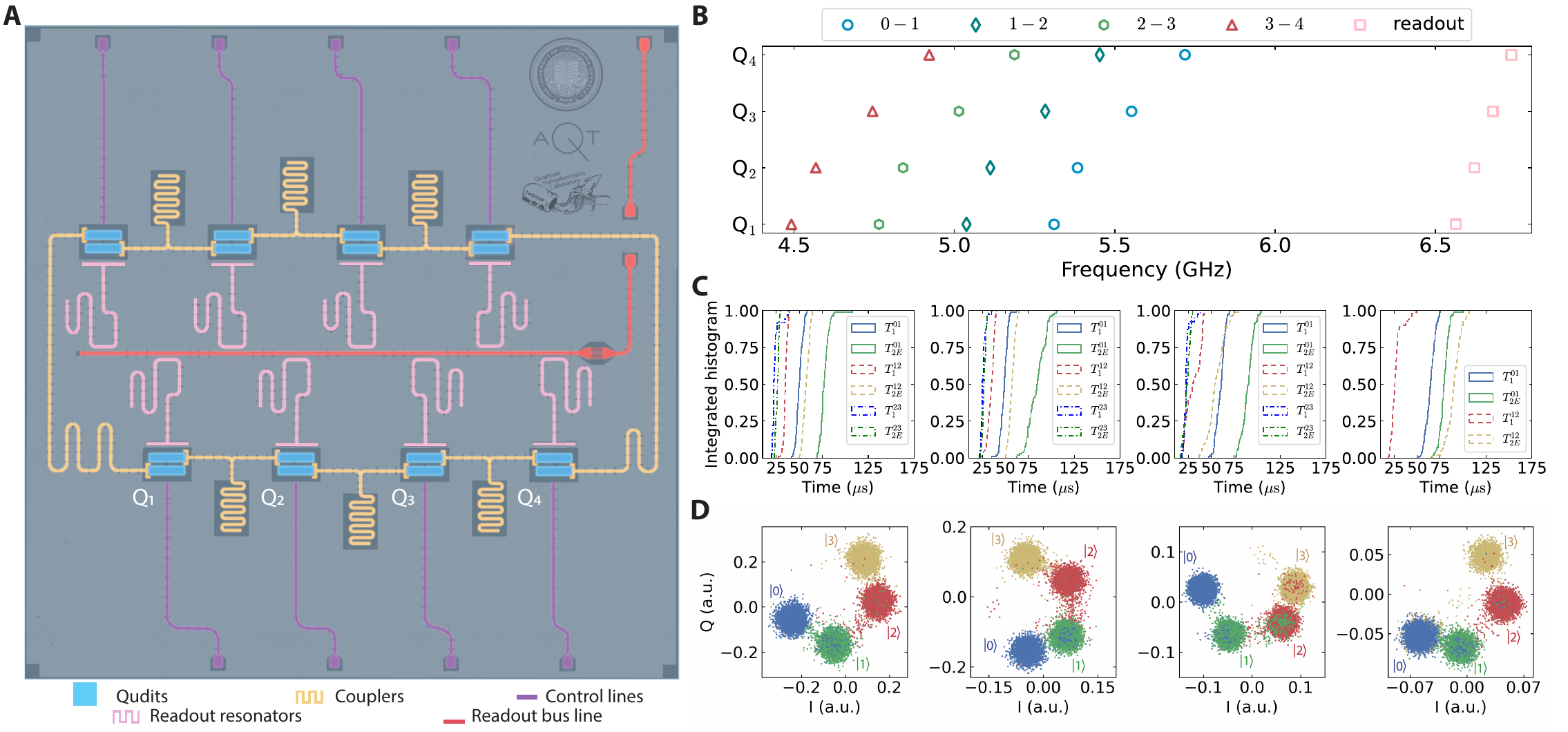}
\caption{\label{figs1} \textbf{Experimental device}. (\textbf{A}) False-color micrograph of the qudit-ring device. The qudits (\textcolor{cyan}{blue}) are transmon circuits. They are pair-wise connected using resonator couplers (\textcolor{yellow}{yellow}). Each qudit is capacitively connected to a microwave control line (\textcolor{purple}{purple}) and dispersively coupled to a readout resonator (\textcolor{pink}{pink}), which are mutually connected to a readout feed line (\textcolor{red}{red}). (\textbf{B}) Frequency allocation of the qudits used in the experiment. (\textbf{C}) Coherence time statistics of the relevant qudit transitions. (\textbf{D}) Dispersive readout histogram of the qudits.}
\end{figure*}
\subsubsection{Device description}
\noindent The experiment is performed on a superconducting device consisting of eight connected transmon circuits forming a ring, as displayed in Fig.~S\ref{figs1}(\textbf{A}), where each transmon serves as the qudit. The capacitive energy is designed to be $E_C\sim 270~\mathrm{MHz}$, and the Josephson energy is used to tune the qudit's $0-1$ frequency to be around $5.5~\mathrm{GHz}$. The qudits are pair-wise connected using resonator couplers. Each qudit is capacitively connected to a separate microwave driving line and dispersively coupled to a readout resonator. The frequency allocation of the qudits and readout resonator used in the experiments is shown in Fig.~S\ref{figs1}(\textbf{B}). The resonator couplers are designed to have their resonant frequencies at approximately $7.2~\mathrm{GHz}$ and qubit-resonator coupling strength of about $75~\mathrm{MHz}$. The effective qudit-qudit subspace coupling can be approximated as
\begin{equation}
    g = \frac{g_{1r} g_{2r}}{2}\left(\frac{1}{\Delta_{1r}} + \frac{1}{\Delta_{2r}}\right),   
\end{equation}
where $g_{ir}$ is the direct coupling coefficient between qudit $i$ and the resonator, $\omega_r$ is the frequency of the resonator, and $\Delta_{ir} = \tilde{\omega}_{q,i} - \omega_r$ is the qudit-resonator detuning. Here, we define $\tilde{\omega}_{q,i} = \omega_{q,i} + g_{ir}^2/(\omega_{q,i} - \omega_r)$, where $\tilde{\omega}_{q,i}$ ($\omega_{q,i}$) is the relevant dressed (bare) transition frequency of qudit $i$. This results in a coupling coefficient $g_{01}/2\pi\sim 3~\mathrm{MHz}$ in the $0-1$ subspace. 

\subsubsection{Device fabrication}
\noindent The device was fabricated on a high-resistivity silicon wafer ($\rho > 10 \text{k}\Omega\text{-cm}$) with Niobium (Nb) and Aluminum (Al). After cleaning the wafer with piranha (mixture of sulfuric acid and hydrogen peroxide heated to $120^\circ \text{C}$) and hydrofluoric acid (HF) to remove the organics and silicon oxide, a 200-nm layer of Nb was sputtered. Then, the superconducting circuit components except the junctions are defined by the photo-lithography technique and reactive-ion etching. We intentionally over-etched the silicon substrate to avoid short circuits and reduce the dielectric loss. 

The wafer was then cleaned with buffered oxide etch (BOE) solvent before the fabrication of Josephson junctions. Josephson junctions are defined by e-beam lithography and deposited by triple-angle e-beam evaporation of Al films following the Manhattan style technique. A gentle plasma cleaning is applied before the Al film deposition to clean the e-beam resist residues after the development and enhance the bond between the substrate and the Al films. The galvanic contact between the Josephson junctions and their capacitor pads was formed by Ar ion-milling band-aid process.

Finally, the fabricated wafer was coated with methyl methacrylate (MMA) resist to protect it before dicing. After dicing, the chips are cleaned with N-methylpyrrolidone (NMP) at $80^\circ \text{C}$ and packaged in a copper box for testing in the dilute refrigerators.

\subsubsection{Single-qudit control}

\begin{figure}
    \centering
    \includegraphics[width = \columnwidth]{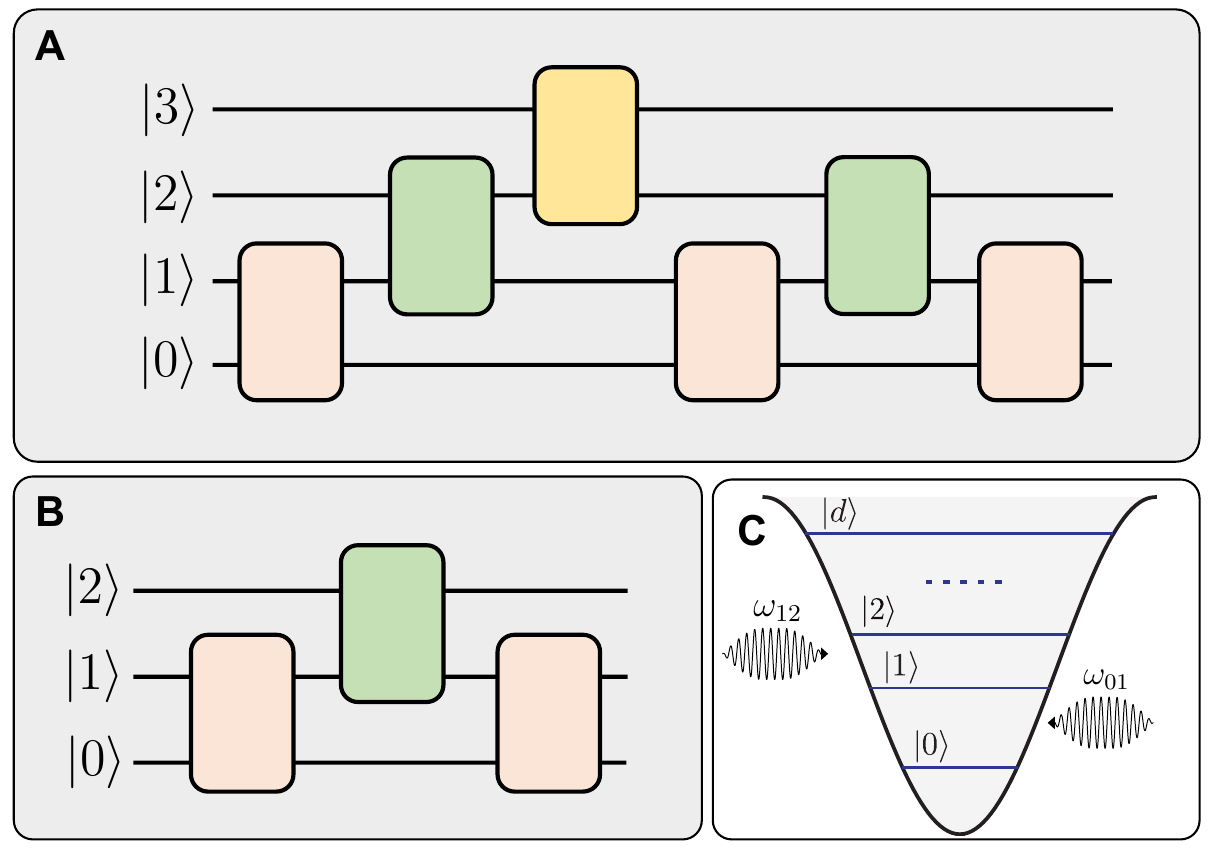}
    \caption{\textbf{Decomposition of control pulses}. (\textbf{A}) An arbitrary SU$(4)$ qudit gate can be decomposed into 6 blocks of SU$(2)$ unitaries in two-level subspaces of the qudit. (\textbf{B}) Similarly, any SU$(3)$ qudit gate can be decomposed using only three subspace SU$(2)$ unitaries. (\textbf{C}) Subspace SU$(2)$ operations are performed via resonant microwave drives at frequency $\omega_{k,k+1}$ and virtual-Z gates.}
    \label{fig:qudit_decomp}
\end{figure}
\noindent The transmon Hamiltonian reads
\begin{equation}
    \hat{H}/h = 4E_C(\hat{n} -n_g) - E_J\cos\hat{\varphi},
    \label{eq:transmon}
\end{equation}
where $\hat{n}$ and $\hat{\varphi}$ are respectively the Cooper-pair number operator and gauge-invariant phase operator. For energy ratio $E_J/E_C>50$, the system admits many bound states that can be spectrally resolved and coherently addressed. One can approximate the Hamiltonian of the transmon as that of a Duffing Oscillator with frequency $\omega_{01}/2\pi \approx \sqrt{8E_JE_C}-E_C$ and self-Kerr $\alpha/2\pi \approx E_C$. By introducing a time-dependent driving term $\epsilon(t)$, we have
\begin{equation}
    \hat{H}/\hbar = \omega_{01} \hat{a}^\dag \hat{a} - \frac{\alpha}{2} \hat{a}^\dag \hat{a}^\dag \hat{a}\hat{a} +  \epsilon(t)(\hat{a}^\dag + \hat{a}),
\end{equation}
where $\hat{a}^\dag(\hat{a})$ is the bosonic creation (annihilation) operator. In this approximation, the relevant single-photon transition frequencies are $\omega_{k,k+1} = \omega_{01} - k\alpha$. By choosing the driving term $\epsilon(t) = \Omega \cos(\omega_d t)$ with driving frequency $\omega_d$ on-resonant with one of the resolvable single photon modes $\omega_{k,k+1}$, we can induce $\ket{k} \leftrightarrow\ket{k+1}$ Rabi oscillations with Rabi rates $\Omega_{k,k+1} = \sqrt{k+1}\Omega$. The characteristics of the enhanced Rabi rate of the higher transitions can be observed in Fig. 1\textbf{B} in the main text. 

The Rabi interactions naturally generate two-level subspace Pauli-X operations. To realize universal single qudit control and perform Z-like operations, we employ virtual-Z gates which update the phase of the relevant carrier waves at $\omega_{k,k+1}$~\cite{PhysRevA.96.022330}. Considering the subspace manifold consisting of $d$ eigenstates in the transmon, we express these native Z operations for a phase update of $\phi$ as,
\begin{equation}
    Z^{k,k+1}(\phi) = \sum_{l=k+1}^{d-1} e^{i \phi} \ket{l}\bra{l} + \sum_{l=0}^{k} \ket{l}\bra{l}
\end{equation}
where $Z^{k,k+1}$ corresponds to an update to the carrier wave at frequency $\omega_{k,k+1}$.

\subsubsection{Compilation of single-qudit gates}
To compile arbitrary single-qudit gates over SU$(d)$, we need to tune up only subspace $X90$ pulses corresponding to $\sqrt{X^{k,k+1}}$, where $X^{k,k+1}$ is the two-level embedded qubit Pauli-X operator (or Gell-Mann operator) $X^{k,k+1} = \ket{k}\bra{k+1} + \ket{k+1}\bra{k}$ for $k\in \mathbb{Z}_{d-1}$. In the case of $d=2$, any SU$(2)$ gate can be compiled from two $X90$ physical pulses via the ZXZXZ decomposition~\cite{PhysRevA.96.022330}. As represented in Fig.~S\ref{fig:qudit_decomp}, from the SU$(2)$ blocks, we can construct SU$(d)$ utilizing 3(6) blocks or 6(12) physical pulses per unitary for $d=3(4)$~\cite{PhysRevA.97.022328}. 

\subsubsection{Qudit dispersive readout}
\begin{figure}
\includegraphics[width=0.45\textwidth]{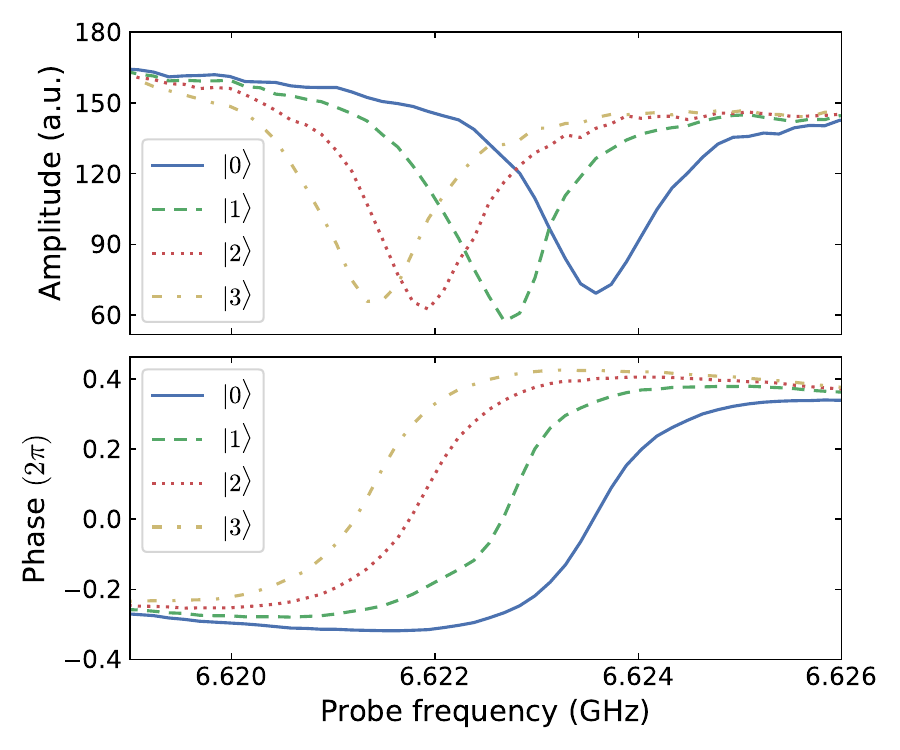}
 \caption{\textbf{Resonator readout signal}. (Top) Magnitude and (Bottom) phase of the resonator probe signal when the qudit is prepared in $\ket{0}$, $\ket{1}$, $\ket{2}$, and $\ket{3}$. 
 \label{fig:res_signal}
}
\end{figure}

\noindent To measure all qudit states, we couple the transmons dispersively to individual resonator modes. This system can be well approximated by the linear dispersive Hamiltonian given in Ref.~\cite{PhysRevLett.105.223601},
\begin{equation}
\begin{split}
    \hat{H}_{JC}^D/ \hbar =& \omega_{r} \hat{b}^\dag \hat{b} + \sum_{k=0}^{d}\omega_k \ket{k}\bra{k} + \sum_{k=1}^{d-1} \chi_{k-1}\ket{k}\bra{k}  \\ - &\sum_{k=0}^{d-1}(\chi_k-\chi_{k-1})\ket{k}\bra{k} \hat{b}^\dag \hat{b},
\end{split}\label{eq:dispersive}
\end{equation}
where $\hat{b}^\dag (\hat{b})$ is the bosonic creation (annihilation) operator associated with the resonator mode at frequency $\omega_r$. Here, we express the Hamiltonian for $d$ levels of the qudit in its own eigenbasis. Following Eq.~\ref{eq:dispersive}, the resonator experiences a dispersive shift of $\chi_{k}-\chi_{k-1}$ when the qudit is in state $\ket{k}$, where $\chi_{k}=g_k^2/\Delta_{kr}$. Figure~S\ref{fig:res_signal} shows the phase and amplitude responses of a resonator mode when a dispersively coupled qudit is prepared in  $\ket{0},\ket{1},\ket{2}$ and $\ket{3}$. For the given $\chi_k$'s and resonator linewidths on our device (varying from 1 to 2 MHz), it is possible to choose a probe frequency for each resonator that distinguishes all four relevant qudit states with single shot separability (see Fig. S\ref{figs1}\textbf{D}).

% \subsubsection{Readout optimization}
% \textbf{\textcolor{blue}{Noah}}

\subsection{Supplementary Note 2 --
Theoretical Analysis\label{sec:SM_theory}}

\noindent We describe here the theoretical framework for the two-photon transitions demonstrated in this work. The two-qudit two-photon transition between states $\ket{k,l}$ and $\ket{k+1, l+1}$ can be understood as a two-qudit generalization of the standard single-atom Raman transition in a ``ladder" configuration in which a single-photon transition from the ground state to the second excited state is not possible, but a two-photon transition mediated by a ``virtual" energy level is possible. Crucially, this transition avoids populating the first excited state while achieving full population transfer between the ground and second excited states.
\begin{figure}
\includegraphics[width=0.48\textwidth]{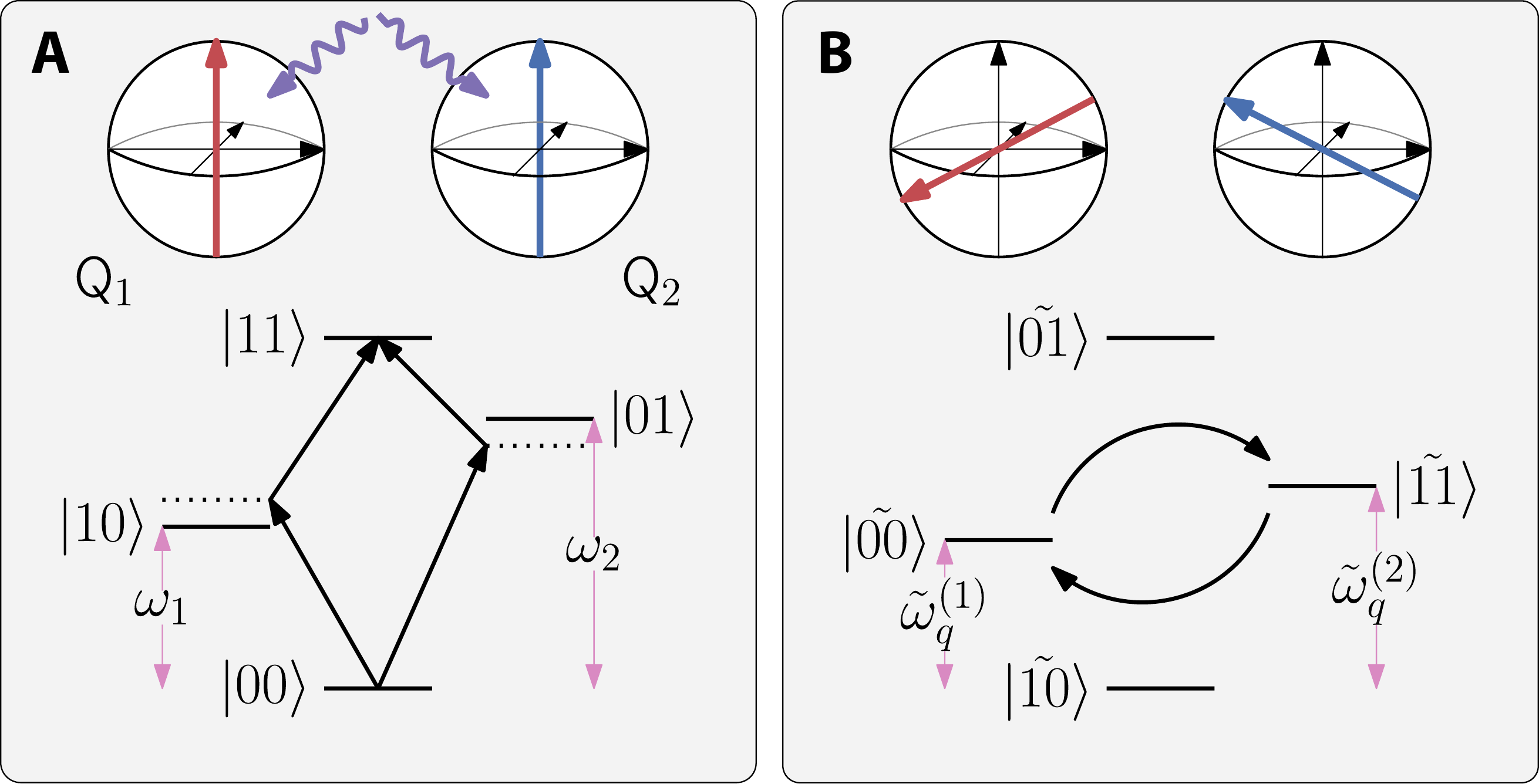}
 \caption{\textbf{Two-photon interaction schematic}. (\textbf{A}) Level diagram of the two-qubit subspace in the laboratory frame. The two-photon transition between $\ket{00}$ and $\ket{11}$ is enabled by a Raman-like mechanism that avoids transitions outside of the computational subspace. (\textbf{B}) Level diagram of the two-qubit subspace in the rotating frame. Although in the laboratory frame, the bare states $\ket{00}$ and $\ket{11}$ appear far from resonance, in the rotating frame, the energies of the corresponding \textit{dressed} states can be tuned into resonance. 
 \label{fig:s2_1}
}
\end{figure}
\subsubsection{Theory of coupled two-level systems}
\noindent We begin by considering two coupled transmons within the two-level system approximation, i.e. treating them as qubits. We presently derive resonance conditions and transition rates for coherent oscillation between the $\ket{00}$ and $\ket{11}$ states. The static Hamiltonian $\Hss$ for this system is given as

\begin{equation}
\begin{split}
    & \Hss = \sum_{i=1}^2\Hq^{(i)}  + \Hc, \\
    & \Hq^{(i)}/\hbar = -\frac{1}{2}\omega_q^{(i)} \hat{\sigma}_z^{(i)}, \\
    & \Hc/\hbar = g \hat{\sigma}_x^{(1)} \hat{\sigma}_x^{(2)},
\end{split}
    \label{eq:Hss_lab}    
\end{equation}
where $\omega_{q}^{(i)}$ (used interchangeably with $\omega_{q,i}$) is the transition frequency between the ground state and the first excited state of the $i$-th qubit and $g$ is the direct qubit-qubit coupling. Without loss of generality, we will work with the assumption $\omega_{q}^{(1) }< \omega_{q}^{(2)}$.

We first consider driving both qubits at the same frequency $\omega_d$, generating the Hamiltonian in the laboratory frame as

\begin{equation}
\begin{split}
    &\Hfull = \Hss + \sum_{i=1}^2 \Hd^{(i)}, \\
    &\Hd^{(i)}/\hbar= \Omega_i \left(\hat{\sigma}_+^{(i)} + \hat{\sigma}_-^{(i)}\right) \cos(\omega_d t + \phi_i), 
\end{split}
    \label{eq:Hfull_lab}
\end{equation}
where $\Omega_i$ is the amplitude of the drive on qubit $i$ (expressed in terms of the resonant Rabi rate) and $\phi$ is the relative phase difference between the drives. We also find it convenient to express the results in terms of the ``drive ratio" $\lambda \equiv \Omega_1 / \Omega_2$. The setup in the lab frame is sketched in  Fig.~S\ref{fig:s2_1}\textbf{A}, forming a ladder-like Raman configuration in which the direct transition between the $\ket{00}$ and $\ket{11}$ states is forbidden but can be mediated by virtual levels~\cite{sorensen1999multiparticle,nesterov2021proposal}. 

The formation of the virtual energy levels is better understood in the rotating frame in which the \textit{dressed} energy levels can be tuned to achieve the desired resonance. In the rotating frame, the Hamiltonians read

\begin{equation}
\begin{split}
    &\Hrotfull = \sum_{i=1}^2 \Hrotqd^{(i)} + \Hrotc, \\
    &\Hrotqd^{(i)} / \hbar = -\frac{1}{2} \left(\Delta_i\hat{\sigma}_z^{(i)} + \Omega_i (\hat{\sigma}_+^{(i)}e^{-i\phi_i} + 
     \hat{\sigma}_-^{(i)}e^{i\phi_i})\right), \\
     &\Hrotc/\hbar = g(\hat{\sigma}_+^{(1)}\hat{\sigma}_-^{(2)} + \hat{\sigma}_-^{(1)}\hat{\sigma}_+^{(2)}),    
\end{split}
    \label{eq:Hfull_rotating}
\end{equation}
where the fast-rotating terms have been omitted. Suppose that at $t=0$, $\Omega_i(t=0) = 0$ and is slowly ramped up to some final value $\Omega_i(t=t_\mathrm{ramp}) = \Omega_i$. As in Ref.~\cite{nguyen2022programmable}, the eigenstates of the uncoupled, undriven Hamiltonians evolve into those of $\Hrotqd^{(i)}$ and are tilted with respect to the $z$-axis of the Bloch spheres. Defining angles $\theta_i$ via
\begin{align}
\begin{split}
    \cos \theta_i &= \frac{\Delta_i}{\sqrt{\Delta_i^2 + \Omega_i^2}}, \\
    \sin \theta_i &= \frac{\Omega_i}{\sqrt{\Delta_i^2 + \Omega_i^2}}, \\    
\end{split}
    \label{eq:angle_def}
\end{align}
the eigenstates of $\Hrotqd$ are given by
\begin{align}
\begin{split}
    \ket{\theta_i} &= \cos\left(\frac{\theta_i}{2}\right) \ket{0} + e^{i\phi_i} \sin\left(\frac{\theta_i}{2}\right) \ket{1}\\
    \ket{\bar{\theta}_i} &= -e^{-i\phi_i}\sin\left(\frac{\theta_i}{2}\right) \ket{0} + \cos\left(\frac{\theta_i}{2}\right) \ket{1}
\end{split}
\end{align}
When $\Delta_i > 0$ ($\Delta_i < 0)$ and $\Omega_i(t)$ is ramped up adiabatically from zero, the state $\ket{0}$ evolves into $\ket{\theta}$ ($\ket{\bar{\theta}}$). The dressed qubit frequency evolves to $\tilde{\omega}_q^{(i)}$, which is given by 
\begin{equation}
   \tilde{\omega}_q^{(i)} = \sgn(\Delta_i) \sqrt{\Delta_i^2 + \Omega_i^2}.
   \label{eq:dressed_qubit_frequency}
\end{equation}
By choosing $\omega_{q}^{(1)} < \omega_d < \omega_{q}^{(2)}$, we achieve the level structure shown in Fig.~S\ref{fig:s2_1}\textbf{B}. Then, by adjusting $\Omega_i$, we seek degeneracy between the levels for states $\ket{\bar{\theta }\theta}$ and $\ket{\theta\bar{\theta}}$. Solving for this resonance condition, we obtain the following relationship between the resonant drive frequency, the drive amplitude, the drive ratio, and the qubit frequencies,
\begin{equation}
    \omega_d = \frac{\omega_q^{(12)}}{2} + \frac{(\lambda^2 - 1)\Omega^2}{2\Delta_q^{(12)}},
    \label{eq:2LS_optimal_frequency}
\end{equation}
where $\omega_q^{(12)} = \omega_q^{(1)} + \omega_q^{(2)}$ and $\Delta_q^{(12)} = \omega_q^{(1)} - \omega_q^{(2)}$.

\begin{figure*}
\includegraphics[width=0.92\textwidth]{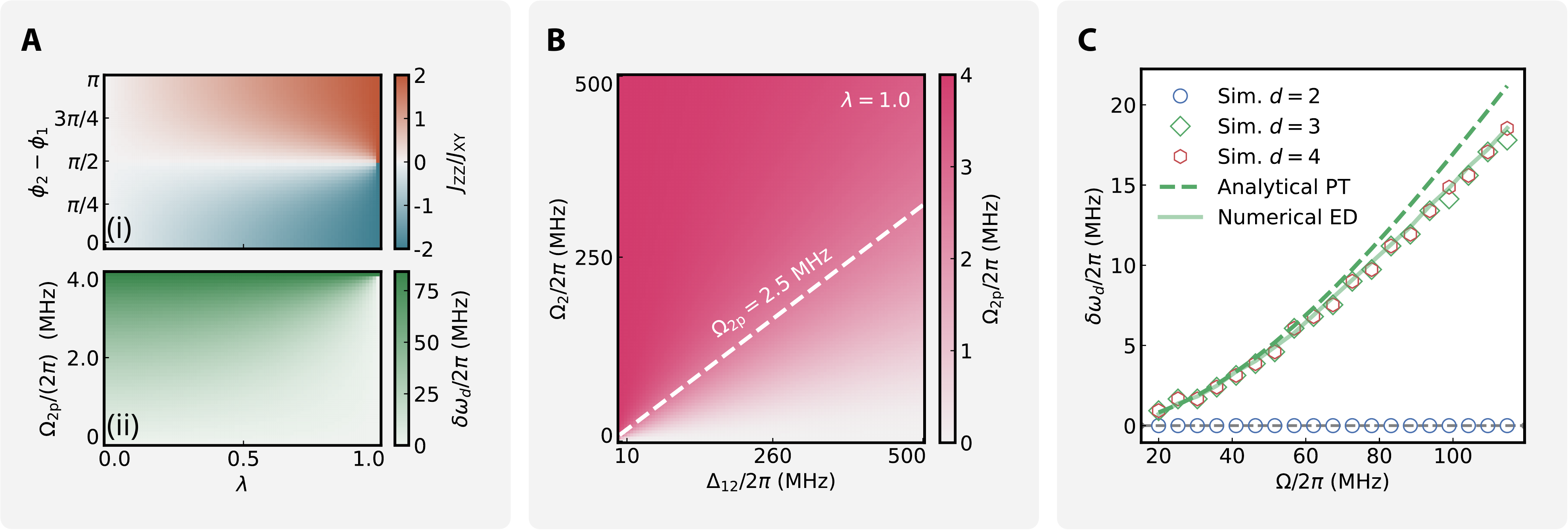}
    \caption{\textbf{Theoretical results}. (\textbf{A}) Tunability of the Hamiltonian through drive ratio and phase. (i) The drive ratio $\lambda = \Omega_1 / \Omega_2$ and relative microwave phase $\phi_2 - \phi_1$ of the single qubit drives can be tuned to apply a two-qubit entangling interaction with a variety of relative interaction strengths $J_\mathrm{ZZ} / J_\mathrm{XY}$ (ii) The two-photon interaction through monochromatic driving can be activated across the range of frequencies between $\omega_{q,1}$ and $\omega_{q,2}$. For simplicity, we have assumed $\Omega_1 \leq \Omega_2$, so $\omega_d \geq (\omega_{q,1} + \omega_{q,2}) / 2$, but the opposite detuning can be achieved by driving qudit 1 harder than qudit 2. (\textbf{B}) Two-photon swap rate $\Omega_\mathrm{2p}$ as a function of the drive amplitude and qudit detuning $\Delta_{12}$. Even when the energy levels $\ket{k+1,l}$ and $\ket{k,l+1}$ are far-detuned, it is possible to activate the two-photon swap, albeit requiring more drive power. (\textbf{C}) Effect of finite anharmonicity on optimal drive frequency. Stark effects from higher levels of the transmon, particularly the third level, shift the dressed qubit frequency and consequently the optimal drive frequency to achieve two-photon transitions. The optimal drive frequency is normalized to the median qubit frequency $\delta \omega_d = \omega_d - (\omega_{q,1} + \omega_{q,2})/2$. Numerical simulations (labeled as ``Sim.") were carried out with $d=\{2,3,4\}$ levels for each transmon to identify the optimal drive frequencies and compared to those found by the exact diagonalization (labeled as ``Numerical ED") of the rotating frame Hamiltonian and analytical perturbation theory (labeled as ``Analytical PT").
    \label{fig:s2_2}
    }
\end{figure*}

Having identified the resonant condition, we enter the frame of $\Hrotqd$. The interaction term $\Hrotc$ is transformed into 
\begin{widetext}
\begin{align}
\begin{split}
    \tilde{H}_c/\hbar &= J_I\left(\tilde{\hat{\sigma}}_x^{(1)} \tilde{\hat{\sigma}}_x^{(2)} - \tilde{\hat{\sigma}}_y^{(1)} \tilde{\hat{\sigma}}_y^{(2)} \right) + J_Q \left(\tilde{\hat{\sigma}}_x^{(1)} \tilde{\hat{\sigma}}_y^{(2)} + \tilde{\hat{\sigma}}_y^{(1)} \tilde{\hat{\sigma}}_x^{(2)} \right) + J_\mathrm{ZZ} \tilde{\hat{\sigma}}_z^{(1)} \tilde{\hat{\sigma}}_z^{(2)},\\ 
    J_I  &= \frac{g}{8}\bigg[\left(1 + \cos\theta_1\right)\left(1+\cos\theta_2\right)\cos 2\phi_1 + \left(1 - \cos\theta_1\right)\left(1-\cos\theta_2\right)\cos 2\phi_2 \bigg], \\
    J_Q &= -\frac{g}{8}\bigg[\left(1 + \cos\theta_1\right)\left(1+\cos\theta_2\right)\sin 2\phi_1 + \left(1 - \cos\theta_1\right)\left(1-\cos\theta_2\right)\sin 2\phi_2 \bigg], \\
    J_\mathrm{ZZ} &= -\frac{g}{2}\sin(\theta_1)\sin(\theta_2) \cos(\phi_1 - \phi_2).   
\end{split}
    \label{eq:2LS_rates}
\end{align}    
\end{widetext}
As expected, the maximal oscillation rate $\Omega_\mathrm{2p}$ in the $00-11$ subspace, $4\sqrt{J_I^2 + J_Q^2}$ is set by the bare coupling $g$. This asymptotic behavior is well-captured by this theoretical approach, which, in contrast to prior work, treats the coupling term $\Hc$ \textit{after} exactly diagonalizing the full uncoupled rotating Hamiltonian $\Hrotqd$ rather than diagonalizing the full, coupled static Hamiltonian first and then treating the driving terms perturbatively. However, standard expressions in the literature, which typically contain terms up to second order in the perturbative parameter $\Omega_i / \Delta_i$, can be recovered by Taylor expansion of Eqs.~\eqref{eq:2LS_rates}. This alternative approach is justified due to the unique hierarchy of energy scales that arises from entangling qubits via strong off-resonant drives: the parameter $\Omega_i / \Delta_i$ is no longer ``small", so corrections to the perturbation theory are significant.

To build intuition for the expressions in Eqs.~\eqref{eq:2LS_optimal_frequency} and ~\eqref{eq:2LS_rates}, we consider two limiting cases. First, when the qubits are driven symmetrically $(\lambda=1)$ and in-phase with the same amplitude $\Omega$, the rates are simplified to
\begin{align}
\begin{split}
    \omega_d &= \frac{\omega_q^{(12)}}{2} \\
    J_I &= \frac{g}{4} \frac{\Omega^2}{\Omega^2 + (\Delta_q^{(12)})^2/4}, \\
    J_Q &= 0, \\
    J_\mathrm{ZZ} &= -\frac{g}{2} \frac{\Omega^2}{\Omega^2 + (\Delta_q^{(12)})^2/4}.
\end{split}
    \label{eq:symmetric_driving}
\end{align}
On the other hand, when only one qubit is driven (e.g. $\lambda = 0$) as done in Ref.~\cite{wei2023native}, we obtain
\begin{align}
\begin{split}
    \omega_d &= \frac{\omega_q^{(12)}}{2} - \frac{\Omega^2}{2\Delta_q^{(12)}}, \\
    J_I &= \frac{g}{4}\left(1 - \frac{\Delta_q^{(12)}/2 - \Omega^2 / \Delta_q^{(12)}}{\sqrt{\Omega^2 + (\Delta_q^{(12)}/2-\Omega^2/\Delta_q^{(12)})^2}}\right) \\
    J_Q &= 0, \\
    J_\mathrm{ZZ} &= 0.
\end{split}
    \label{eq:one_qubit_driving}
\end{align}
Eqs.~\eqref{eq:2LS_optimal_frequency} and \eqref{eq:2LS_rates} suggest that by controlling the drive ratio $\lambda$ and the relative phase of the microwave drives on each transmon, $\phi=\phi_2 - \phi_1$, one can control (i) the ratio of the two-photon excitation term $J_\mathrm{XY} = \sqrt{J_I^2 + J_Q^2}$ to the ZZ term $J_\mathrm{ZZ}$, as well as (ii) the optimal drive frequency $\omega_d$. We demonstrate the flexibility of this control in Fig.~S\ref{fig:s2_2}\textbf{A}. We speculate that this control could be useful for performing quantum simulations of many-body systems with nearest-neighbor couplings using fewer two-qubit gates, as one can tune ratios of the interaction terms in the Hamiltonian easily.

Our analysis has thus far concentrated on the two-photon transition between the $\ket{00}$ and $\ket{11}$ states of a system of two coupled qubits. However, we can apply a similar analysis to subspaces consisting of higher levels to obtain rates for general high-dimensional systems, assuming that all other levels are far enough apart that they do not contribute significantly to the dynamics. To obtain the expressions for the $\ket{k,l}\rightarrow \ket{k+1, l+1}$ transition, the detunings $\Delta_i$ and $\Delta_q^{(12)}$ are now defined relative to the $k\rightarrow (k+1)$ and $l \rightarrow (l+1)$ transition frequencies of the qudit pair. The drive amplitude $\Omega_1$ ($\Omega_2$) is defined with respect to the Rabi rate in the $\ket{k}, \ket{k+1}$ ($\ket{l}, \ket{l+1}$) subspace when this transition is driven resonantly, and $g$ is replaced by the bare coupling between states $\ket{k,l+1}$ and $\ket{k+1,l}$. The latter typically scales as $\sqrt{(k+1)(l+1)}$ due to the matrix elements of the bosonic ladder operator. Indeed, even when the single-photon transition energies from $\ket{k,l}$ to  $\ket{k+1,l}$ and $\ket{k,l+1}$ are far-detuned, it is still possible to activate the two-photon transition from $\ket{k,l}$ to $\ket{k+1, l+1}$, although a larger drive amplitude (in the relevant qudit subspaces) is probably needed. We illustrate this point in Fig.~S\ref{fig:s2_2}\textbf{B}, which shows the numerically simulated two-photon rate with varying drive amplitudes and detuning. 

\subsubsection{Finite anharmonic correction}
\noindent The preceding analysis of the $\ket{00}\rightarrow \ket{11}$ two-photon transition assumes that states outside the computational subspace are sufficiently off-resonant such that they do not contribute significantly to the dynamics, which is a good assumption for systems with large anharmonicity~\cite{nesterov2021proposal,nguyen2022blueprint}. However, deviations from this theory manifest when higher levels of the transmon are taken into account. The strong off-resonant drive induces a Stark shift of the dressed qubit frequency in Eq.~\eqref{eq:dressed_qubit_frequency}, which must be taken into account when solving for the optimal drive frequency $\omega_d$ at a given drive amplitude $\Omega$. 

We examine this effect by performing numerical simulation of two symmetrically driven transmons using QuTiP~\cite{Johansson2012}, retaining up to four energy levels per transmon as shown in Fig.~S\ref{fig:s2_2}\textbf{C}. As predicted by Eq.~\eqref{eq:2LS_optimal_frequency}, when only two levels are retained, the optimal frequency $\omega_d$ is independent of the drive strength $\Omega$. However, when higher levels are taken into account, a nontrivial dependence manifests. This dependence can be understood by considering the Hamiltonian of the three-level Duffing Oscillator in the rotating frame of the drive for each transmon separately,
\begin{equation}
    \hat{H}_i / \hbar = \left(\Delta_i  - \frac{\alpha_i}{2} \right) \hat{a}_i^\dagger \hat{a}_i + \frac{\alpha_i}{2}(\hat{a}_i^\dagger \hat{a}_i)^2 + \frac{\Omega_i}{2} \left(\hat{a}_i + \hat{a}_i^\dagger\right),
    \label{eq:rotating_Duffing}
\end{equation}
where $\alpha_i$ is the anharmonicity of qudit $i$, $a_i$ is the bosonic annihilation operator for the modes of qudit $i$, and the fast-rotating terms have been dropped. By considering $\Omega_i$ as a parameter which is slowly turned on from $0$ at $t=0$ to their final values $\Omega_i$ at $t=t_\mathrm{ramp}$, one can follow the eigenstates of the rotating frame Hamiltonian from $t=0$ to $t=t_\mathrm{ramp}$ and compute the energy difference $\tilde{\omega}_q^{(i)}$ between the states $\ket{0}$ and $\ket{1}$. This generalizes the dressed qubit frequency introduced in Eq.~\eqref{eq:dressed_qubit_frequency}. Similarly, the resonance between the $\ket{00}$ and $\ket{11}$ states is achieved at a given drive power when $\omega_d$ is tuned such that $\tilde{\omega}_q^{(1)}+\tilde{\omega}_q^{(2)} = 0$. The entire procedure of diagonalization and solution for $\omega_d$ can be carried out numerically or by standard Schrieffer-Wolff perturbation theory to obtain expressions for $\{\tilde{\omega}_q^{(i)}\}$ and then again using numerics to solve for resonance. We compare these two approaches to the optimal frequency obtained by simulations of spectroscopy in QuTiP~\cite{Johansson2012} in Fig.~S\ref{fig:s2_2}\textbf{C}. The agreement between the numerical simulation and the semi-analytical approaches for $d > 2$ suggests that the nontrivial dependence of $\omega_d$ on $\Omega$ can indeed be understood through the framework of effective transition frequencies used in this work.

\subsection{Supplementary Note 3 --
Extended Two-photon Interaction Data}

\begin{figure*}[t]
    \includegraphics[width=0.8\textwidth]{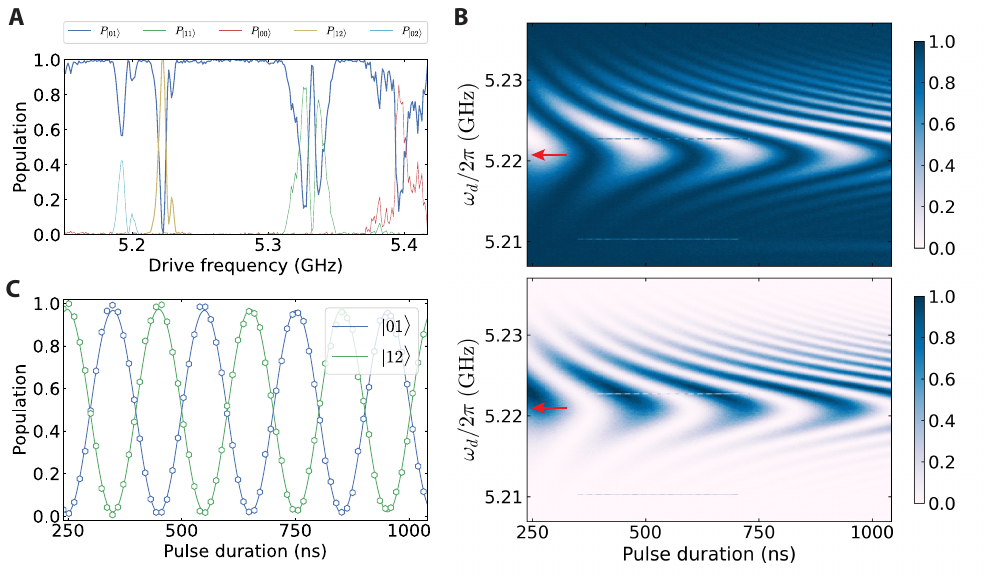}
\caption{\label{figs4} \textbf{$|01\rangle \leftrightarrow |12\rangle$ interaction}. (\textbf{A}) Transition spectrum of a pair of qudits prepared in $|01\rangle$, revealing the sign of the transition. (\textbf{B}) Chevron pattern of the transition with varying drive frequency and pulse duration. (\textbf{C}) Coherent driven oscillation between $|01\rangle$ and $|12\rangle$ originating from the two-photon dynamics. The drive frequency is indicated by the red arrows in panel \textbf{B}.}
\end{figure*}

\noindent The two-photon interaction can also be realized between $\ket{k,l}$ and $\ket{k+1,l+1}$ for $k\neq l$. To provide the supporting data, we showcase the microwave-driven exchange interaction between $|01\rangle$ and $|12\rangle$. The first evidence appears in a spectroscopy measurement in which we prepare the qudits to be in $|01\rangle$ and then sweep the drive frequency across a wide range, as shown in Fig.~S\ref{figs4}\textbf{A}. Using the single-shot readout capability, we can deterministically compute the probability of the qudits in specific states after applying the drives at each frequency point.

The coherent population transfer is then further verified by performing a time-domain measurement of the qudit population after applying pulses with varying durations. As the drive frequency and duration are swept across the preliminary data point in the spectral domain, the exchange interaction manifests as a chevron pattern in Fig.~S\ref{figs4}\textbf{B}. Fixing the drive frequency at the optimal (symmetric) point, we can inspect the microwave-induced oscillation between $|01\rangle$ and $|12\rangle$, as shown in Fig.~S\ref{figs4}\textbf{C}.
\subsection{Supplementary Note 4 --
Extended Data \\ for CnZ gates}

\begin{figure}[t]
    \includegraphics[width=0.5\textwidth]{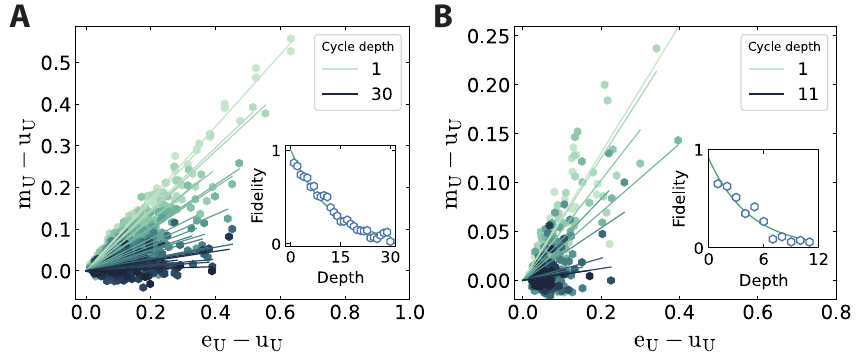}
\caption{\label{figs5} \textbf{Cross-entropy benchmarking}. (\textbf{A}) CCZ result. The differences in the normalized linear cross-entropies for ideal ($m_U-u_U$) and measured ($e_U-u_U$) are shown, which can be used to extract the cycle fidelity. Varying the cycle depth results in an exponential decay (inset) that embodies the dressed cycle fidelity. (\textbf{B}) CCCZ result. The increased cycle error cuts down the maximum cycle depth required.}
\end{figure}

\noindent We leverage Cross-Entropy Benchmarking (XEB)~\cite{arute2019quantum} to further characterize the multi-qubit operation. The cross-entropy $H(p_1,p_2)$ is a statistical measure of the similarity between two probability distributions $p_1$ and $p_2$. In characterizing the fidelity of a quantum operation, we consider the probability distribution $p(x)$ giving the probability of measuring output bitstring $x$ after applying the operation.

Let us define $p$ as the ideal probability, $q$ as the experimental one, then the XEB circuit fidelity is given as 
\begin{equation}
    \mathcal{F}_\mathrm{XEB} = \frac{H(p,q) - H(p,u)}{H(p,p) - H(p,u)}\equiv \frac{m_U-u_U}{e_U-u_U},
\end{equation}
where $H(p_1,p_2) = \sum_x p(x)p_2(x)$ is the linear cross-entropy between two probability distributions $p_1(x)$ and $p_2(x)$, and $u(x)=1/d$ is the uniform probability distribution on the bitstrings. The concept of fidelity in this context can be interpreted as the difference in the ideal to measured and ideal to uniform cross entropies, normalized by the difference if the measured distribution was to perfectly match the ideal distribution. 

For a randomized circuit, the error can be viewed as the deviation of the measured bitstring distributions from a uniform distribution. Notably, XEB does not involve finding the inverse gate. Therefore, it is a versatile tool to characterize non-Clifford and multi-qubit gates such as those in the CnZ gate family. By interleaving the implemented gate with $\mathrm{SU}(2)$ local rotations, we can tailor the gate errors into a global depolarizing channel.

Performing XEB measurement, we obtain the dressed cycle fidelity at different cycle depths. Fitting this to the simple depolarizing noise model gives us the average fidelity at every depth. By varying the depth, we can extract the average cycle fidelity, as shown in Fig.~S\ref{figs5}. The dressed cycle fidelity for the CCZ gate is 91.9(3)\%, approximately equal to the dressed cycle fidelity given by the CB method (92.44(6)\%), albeit with higher uncertainty. This implies that we can roughly estimate the gate fidelity by combining the XEB fidelity with the CB reference cycle fidelity. This approach allows us to estimate the CCCZ gate fidelity to be 88(2)\%, lower than the truth table fidelity. This implies a finite correlated phase error in the obtained result. We note that our numerical simulation shows that the small phase shift of the target qubit associated with the control state $|111\rangle_c$ accounts for an error of less than 0.5\%.

\subsection{Supplementary Note 5 --
Quantum Tomography \\ of High-dimensional States}

\begin{figure*}[t]
    \centering
    \includegraphics[width=0.95\textwidth]{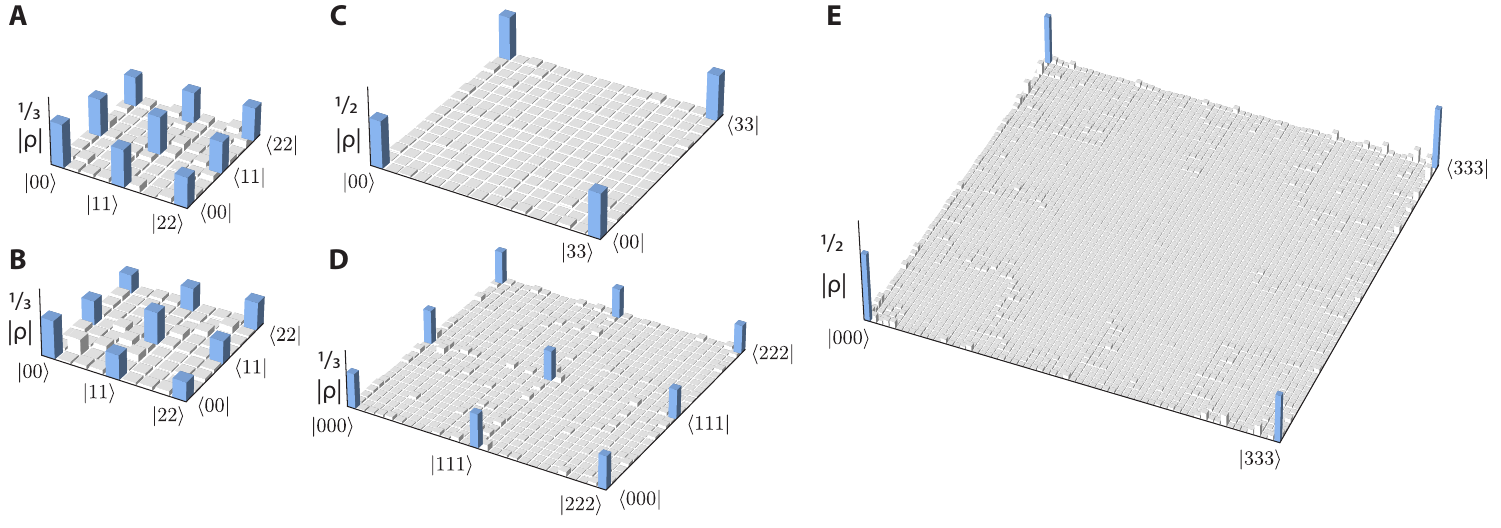}
    \caption{\textbf{Extended tomography results}. (\textbf{A}) Q$_1$-Q$_3$ qutrit Bell state. The raw (purified) fidelity is 75.7\%~(96.3\%). (\textbf{B}) Q$_1$-Q$_4$ qutrit Bell state with $\mathcal{F}= 53.0\%~(95.7\%)$. (\textbf{C}) Schrödinger cat state $(|00\rangle + |33\rangle)/\sqrt{2}$ with $\mathcal{F}= 98.6\%~(99.3\%)$. (\textbf{D}) High-dimensional GHZ state with $n=d=3$ and $\mathcal{F}=90.8\%~(96.4\%)$. (\textbf{E}) Schrödinger cat state $(|000\rangle + |333\rangle)/\sqrt{2}$ with $\mathcal{F}=80.1\%~(90.9\%)$.}
    \label{fig:e_tomo}
\end{figure*}

\noindent To characterize the fidelity of the realized quantum states, we utilize quantum state tomography which has been previously applied to isolated~\cite{PhysRevLett.105.223601, PhysRevX.13.021028, cao2023emulating} as well as multipartite qudit states~\cite{goss2023extending, blok2021quantum, cervera2022experimental}. To reconstruct an arbitrary single~qudit state, we perform projections over an informationally complete set of local rotations. To limit the effects of the state preparation and measurement (SPAM) errors, we choose these to be closest to the native gate set of our device. We note that the Gell-Mann operators, given by $X^{k,l} = \ket{k}\bra{l} + \ket{l}\bra{k}$ and $Y^{k,l} = i(\ket{k}\bra{l} - \ket{l}\bra{k})$, correspond to the native rotations induced by Rabi oscillations at $\omega_{k,k+1}$ for the case of $l=k+1$. We thus choose the natural set of local projections for state tomography as projections onto all the computational states $\ket{0},\dots \ket{d-1}$ as well as two-level $\sqrt{X^{k,k+1}}$ and $\sqrt{Y^{k,k+1}}$ projections within each subspace, or explicitly, the set of projectors $M$ given by
\begin{equation}
\begin{split}
    M = &\{ I, X^{0,1}, X^{1,2}X^{0,1}, X^{2,3}X^{1,2}X^{0,1}, \sqrt{X^{0,1}}, \sqrt{Y^{0,1}},\\ &\sqrt{X^{1,2}}X^{0,1}, \sqrt{Y^{1,2}}X^{0,1}, X^{1,2}\sqrt{X^{0,1}}, X^{1,2}\sqrt{Y^{0,1}}, \\
    &X^{2,3}X^{1,2}\sqrt{X^{0,1}}, X^{2,3}X^{1,2}\sqrt{Y^{0,1}}, X^{2,3}\sqrt{X^{1,2}}X^{0,1}, \\
    &X^{2,3}\sqrt{Y^{1,2}}X^{0,1}, \sqrt{X^{2,3}}X^{1,2}X^{0,1}, \sqrt{Y^{2,3}}X^{1,2}X^{0,1}   \}.
\end{split}
\end{equation}
For the projections of an $n$-qudit state, we simply apply the projections in $M_n = M^{\otimes n}$.

To further mitigate the effects of SPAM measurements, we characterize the readout confusion matrix (or readout misassignment matrix see Fig.~1\textbf{B}) over the relevant $n$ qudit states $\{ \ket{0} \dots \ket{d-1} \}^{\otimes n}$. Given the fidelity of our native operations characterized by RB, we can assume that the readout assignment errors are dominant. This allows us to account for the measurement errors by applying the inverse of the confusion matrix to our averaged ditstring results. Then, the qudit density matrices are reconstructed via maximum likelihood estimation (MLE). In addition to the results presented in Figures~4 and 5 in the main text, the reconstructed density matrices of the non-local qutrit Bell states between Q$_1$-Q$_3$ and Q$_1$-Q$_4$, two and three qudit cat states, and three qutrit GHZ states can all be found in Fig. S\ref{fig:e_tomo}. The state fidelity is computed following the relation
\begin{equation}
    \mathcal{F}( \rho, \sigma) = \text{Tr}\left(\sqrt{\rho} \sigma \sqrt{\rho}\right)
\end{equation}
where $\rho$ is the experimentally reconstructed density matrix, and $\sigma$ is the ideal density  matrix. Finally, we note that the diagonal terms in the density matrix can be measured directly to reduce fluctuation effects arising during the long tomography measurement.

\subsection{Supplementary Note 6~--~\\
Shot Noise Monte Carlo Simulation}

\noindent In this study, the assessment of entangled multidimensional states is conducted through quantum state tomography (QST), a method that reconstructs the density matrices of states using measurements of Pauli operators and MLE. To evaluate the effectiveness of this approach, it is crucial to comprehend the inherent noise in QST and its dependence on the dimensions of the qudits, denoted as $d$, and the number of qudits, denoted as $n$. Specifically, we delve into the impact of shot noise, an inherent element of the QST procedure. This noise arises from the discrepancy between the sample expectation value and the true expectation value, a consequence of a finite number of measurement repetitions (or shots) denoted as $n_\mathrm{rep}$. As per the Central Limit Theorem, the shot noise scales with $1/\sqrt{n_\mathrm{rep}}$ when $n_\mathrm{rep}\gg 1$.

\begin{figure}
\includegraphics[width=0.49\textwidth]{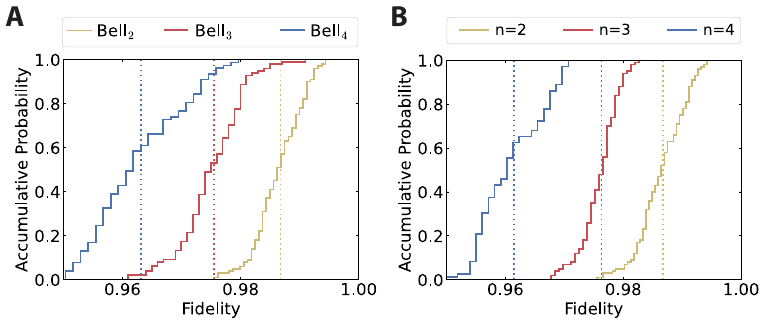}
 \caption{\textbf{Shot noise simulation}. (\textbf{A}) Accumulative probability distribution for qudit Bell states $\text{Bell}_2, \text{Bell}_3, \text{Bell}_4$ with $d = 2, 3, 4$, where the fidelities are estimated to be $0.987(4)$, $0.976(6)$, $0.963(8)$. (\textbf{B}) Accumulative probability distribution for 2 qubit, 3 qubit and 4 qubit GHZ states, where the fidelities are estimated to be $0.987(4)$, $0.976(3)$, $0.96(6)$. \label{fig:s9_1}
}
\end{figure}

First, we investigate the scaling behavior of shot noise with an increasing qudit dimension $d$. This is accomplished by simulating tomography experiments for qudit Bell states $\text{Bell}_2, \text{Bell}_3, \text{Bell}4$ with $d = \{2, 3, 4\}$. The accumulative probability distribution of the fidelities between the QST-reconstructed density matrix and the corresponding Bell state is shown in Fig. \ref{fig:s9_1}(A). The average fidelities are 0.987, 0.976, 0.963, with the standard deviations of 0.004, 0.006, 0.008, respectively, where each distribution consists of 100 simulations of a tomography experiment with $n_\mathrm{rep} = 1000$. The bias of the average fidelities away from 1 is attributed to the MLE process~\cite{silva2017investigating}, while the standard deviation arises from shot noise. Despite the exponential enlargement of the Hilbert space with the increasing qudit dimension $d$, the shot noise is expected to increase only quadratically with $d$ due to the entangling structure of the Bell states~\cite{bavaresco2018measurements}.

Then, we note that increasing the number of qudits also increases the Hilbert space dimension. With a similar method, we estimate the shot noise of the QST for GHZ states involving 2, 3 and 4 qubits ($d=2$) and show their accumulative probability distribution in Fig.~\ref{fig:s9_1}(B). The average fidelities are 0.987, 0.976, 0.961, and the standard deviations are 0.004, 0.003, 0.006. We observe no significant increase or clear scaling relations between the shot noise and the qubit numbers, which can be explained by the entangling structure of the GHZ states, again~\cite{bavaresco2018measurements}. 

This stems from the fact that the fidelity of the Bell states is predominantly determined by the population terms $\langle i\dots i|\rho|i\dots i\rangle$ and the coherence terms $\langle i\dots i|\rho|j\dots j\rangle$, so the uncertainties remain the same for larger systems~\cite{bavaresco2018measurements}. Meanwhile, the number of important terms in the fidelity increases with the number of constituents in the state, which explains the previous observation. We note that our experiment involves entangled states with up to 4 constituents, and thus, the simulation of shot noise for Bell$_d$ with $d=2,3,4$ encompasses the important standard deviations for all the reported results.
\subsection{Supplementary Note 7 --\\
Phase-space Description of High-dimensional States}

\begin{figure*}[t]
    \centering
    \includegraphics[width=0.9\textwidth]{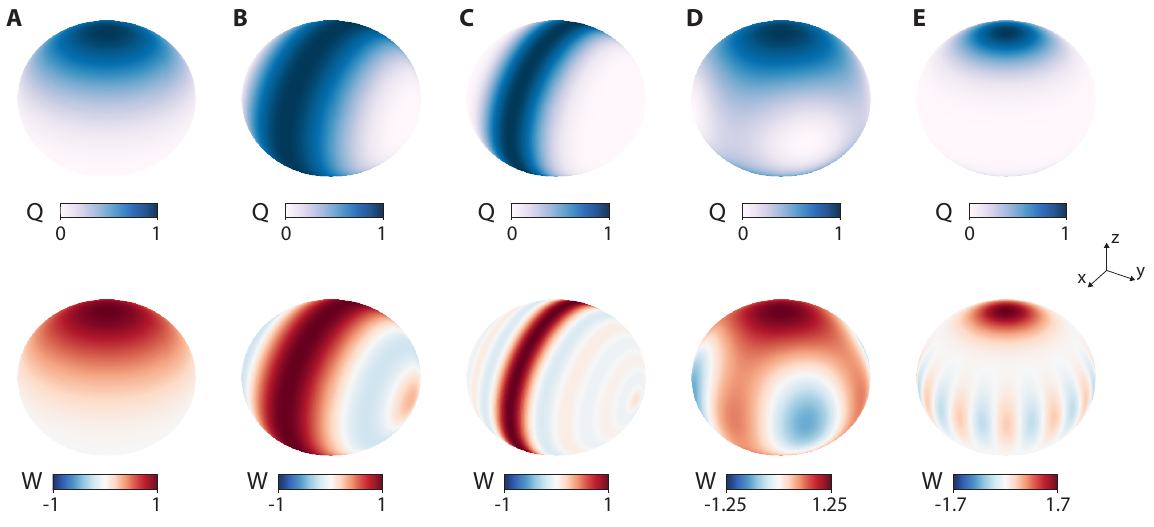}
    \caption{\textbf{Quasiprobability distributions of high-dimensional states}. The top row shows the Husimi-Q distributions. The bottom row shows the Wigner functions. (\textbf{A}) Coherent state $|00\rangle$ with $d=3$. (\textbf{B}) Bell$_3$ state. (\textbf{C}) Bell$_7$ state. (\textbf{D}) Cat$_3$ state with $n=2$. (\textbf{E}) Cat$_7$ state with $n=2$.}
    \label{fig:coherent_states}
\end{figure*}

\noindent In the realm of quantum mechanics, the choice of a representation is often guided by practical considerations. In this vein, the concept of coherent states has emerged as a fundamental framework for standard descriptions within the field of quantum optics, owing to their attractive characteristics: (i) They are evolved from the vacuum state by a unitary operator, and are minimum-uncertainty states. (ii) They obey a completeness relation and thus form a good set of basis states. (iii) They provide a quantum approach to conceptualize classical fields. Notably, macroscopic quantum states of atomic ensembles can also be described by such a concept. This duality is related to the group contraction of spin operators based on the angular momentum algebra to bosonic operators based on the harmonic-oscillator algebra~\cite{arecchi1972atomic}.

A \textit{spin coherent state} (SCS) $|\theta,\phi\rangle$ is defined as an eigenstate of a spin component in the ($\theta,\phi$) direction, $S_{\theta,\phi}=S_x\sin\theta\cos\phi+S_y\sin\theta\sin\phi+S_z\cos\theta$. Here, $\theta$ denotes the polar angle, $\phi$ denotes the azimuthal angle, and the spin system obeys the cyclic commutation relation $[S_i,S_j]=i\epsilon_{ijk}S_k$, where $\epsilon_k$ is the Levi-Civita symbol. 
The associated uncertainty relation is $\langle \Delta S_i ^2\rangle \langle \Delta S_j^2\rangle \geq \frac{1}{4} |\langle S_k\rangle|^2$. An SCS satisfies the minimal uncertainty relation with equal uncertainties in both orthogonal components, corresponding to an isotropic Husimi-Q quasiprobability distribution in the spherical phase space~\cite{arecchi1972atomic,kitagawa1993squeezed}, as shown in Fig.~S\ref{fig:coherent_states}\textbf{A}. In the present framework, we project a high-dimensional spin-$j$ state $|j\rangle$, where $j=\frac{1}{2}(d-1)$, onto an SCS for useful visualization using the generalized Bloch sphere,  
\begin{equation}
    |\theta,\phi\rangle = \exp \left[\frac{1}{2}\theta (e^{i\phi}J_- - e^{-i\phi}J_+ )\right] |j\rangle,
\end{equation}
where $J_\pm = J_x\pm iJ_y$ are the raising/lowering angular momentum operators, and we denote $|j\rangle\equiv |j,m=j\rangle$ for a general spin-$j$ system~\cite{arecchi1972atomic}.

An ensemble of $n$ pure SCS's simply form a product state,
\begin{equation}
    \prod_{k=0}^{n-1}|\theta_k,\phi_k\rangle_k=\prod_{k=0}^{n-1}\left[\cos\frac{\theta_k}{2}|0\rangle_k +\sin \frac{\theta_k}{2}e^{i\phi_k}|1\rangle_k\right],
\end{equation}
where $|0\rangle_k$ and $|1\rangle_k$ are eigenstates of $S_z$ in the $k$th SCS.
If all the individual constituents are aligned along the same direction ($\theta,\phi$), the resulting state becomes the collective SCS, 
\begin{equation}
    |\theta,\phi\rangle =\left[\cos\frac{\theta}{2}|0\rangle +\sin \frac{\theta}{2}e^{i\phi}|1\rangle\right]^{\otimes n}.
\end{equation}
This alignment can be conveniently realized via initialization of all the qudits to the same state. For a collection of spin-$j$ system with dimension $d+1$, if the basis SCS is projected from $|jj\dots j\rangle$ with $j=d/2$, then the high-dimensional states with isotropic QPD are $|00\dots 0\rangle$ and $|dd\dots d\rangle$. 

Starting from the optical coherent state, various interesting states of light can be realized. We presently explore the duality of these states in qudit systems starting from the SCS. An important concept in quantum optics is \textit{squeezing}, which is paradigmatically generated by the operator $\hat{\mathcal{S}}(\zeta)=\exp \left[\frac{1}{2}(\zeta\hat{a}^{\dagger 2}-\zeta^*\hat{a}^2) \right]$, where $\zeta$ is the squeezing parameter and $\hat{a}$ ($\hat{a}^\dagger$) is the bosonic annihilation (creation) operator. 
% This draw analogy with the two-axis twisting Hamiltonian in atomic system which corresponds to the simultaneous excitation-deexcitation of two atoms~\cite{kitagawa1993squeezed}. We note the similarity between the quantum electrodynamics associated with $\hat{\mathcal{S}}$ and the effects induced by our two-photon drive Hamiltonian. 
Specifically, the photonic two-mode squeezing process reads $\hat{\mathcal{S}}(\zeta)=\mathrm{exp}(\zeta\hat{a}^\dagger\hat{b}^\dagger - \zeta^*\hat{a}\hat{b})$, where $\zeta$ is the squeezing strength, and $\hat{a}$ ($\hat{b}$) is the annihilation operator of the first (second) mode. 

The process results in the two-mode squeezed state, $|\Psi\rangle_\mathrm{2ms}\propto \sum_{N=0}^\infty c_N |NN\rangle$, where $c_N$ is a coefficient dependent on $\zeta$ and $N$ is the quanta number~\cite{caves1985new}. In a coupled system comprising of two qudits with dimension $d$, if we draw the duality of $N$ ranging from 0 to $d-1$, the high-dimensional Bell$_d$ state would be close to spin squeezed state (SSS). As opposed to the SCS, the SSS is signified by the shrinking of its Husimi-Q distribution along a geodesic, which is further enhanced by increasing the dimension of the system, as shown in Fig.~S\ref{fig:coherent_states}\textbf{B,C}.

Another important bosonic state of immense interest is the Schrödinger cat state. In atomic systems, this represents a superposition of macroscopic SCS's. The structure of high-dimensional systems and the two-photon dynamics suggest that Schrödinger cat states can be realized efficiently in qudits. We demonstrate this by creating the qudit Schrödinger cat states, which are superpositions of the qudit coherent states, as discussed in the main text. Importantly, the size of the realized cat depends equally on the dimension $d$ and the number of parties $n$. We provide a gallery of Wigner functions for different cat states in  Fig.~S\ref{fig:coherent_states}\textbf{D,E}.

\subsubsection{Quasiprobability distributions}

\noindent The concepts of squeezing and macroscopic superpositions are more conveniently represented using quasiprobability distribution functions in phase space, such as the Husimi-Q function or the Wigner function. In particular, the Husimi-Q distribution is frequently employed to display the macroscopic aspect of entangled quantum systems~\cite{song2019generation}, due to its relatively straightforward calculation procedure via the formula
\begin{equation}
    Q(\theta,\phi)\propto \langle \theta,\phi |\hat{\rho} | \theta,\phi \rangle.
\end{equation}
This involves computing the overlap probability between the target state specified by $\hat{\rho}$ and the SCS $|\theta,\phi \rangle$ as the basis state. The Husimi-Q function is nevertheless not suitable to exhibit the non-classical feature of macroscopic superposition states such as the Schrödinger cats, as showcased in Fig.~S\ref{fig:coherent_states}\textbf{D,E}.

The Wigner function's ability to take on negative values, on the other hand, has proven to be transformative in the visualization of quantum correlations. It has thereby become an indispensible tool in the analysis of cat states. The Wigner function can be viewed as the expectation value of a normalized parity operator, which renders it possible to be represented using different bases. Here, we follow the convention using tensor product kernels prescribed by Ref.~\cite{rundle2017simple},
\begin{equation}
    W_{\hat{\rho}}(\theta,\phi) = \langle \hat{U}(\theta,\phi) \hat{\Pi} \hat{U}^\dagger (\theta,\phi)\rangle _{\hat{\rho}}.
\end{equation}

Here, $\hat{\rho}$ is the density matrix, $\hat{U}$ is the general displacement or rotation operator, and $\hat{\Pi}$ can be viewed as the spin parity operator, analogous to the bosonic case. These operators are not unique, and they only have to obey the Stratonovich-Weyl correspondence. Our capability to measure high-dimensional systems motivates the use of a tensor product of spins. This simply involves laying out the SU$(d)$ rotation of a single-qudit, using $\hat{\Pi}_{\otimes^n \mathrm{SU}(d)}=\otimes^n \hat{\Pi}_{\mathrm{SU}(d)}$ to achieve
\begin{equation}
    W_{\otimes^n \mathrm{SU}(d)} (\theta,\phi) = \mathrm{Tr}\left[\hat{U}(\theta,\phi) \hat{\rho} \hat{U}^\dagger (\theta,\phi) \hat{\Pi}_{\otimes^n \mathrm{SU}(d)}\right].
\end{equation}
Following this convention, we can use the predefined functions in QuTiP~\cite{Johansson2012} to compute the QPD for density matrices of arbitrary spin dimensions. Our analysis involves the normalization of the Husimi-Q QPD values such that the highest probability amplitude is unity. The computed Wigner function is kept as-is due to the physical significance of the values. A gallery of both functions for two-mode qudit states is displayed in Fig.~S\ref{fig:coherent_states}.

 \subsection{Supplementary Note 8 -- \\
High-dimensional Quantum Circuit Synthesis}

\noindent This note describes our approach for synthesizing high-dimensional state preparation circuits. Using this procedure, we developed the analytical solutions for preparing GHZ circuits of any size $n$ and systems of any dimension $d$, using the reported two-photon transitions.

\subsubsection{Two-qudit Bell state preparation}
\noindent We begin by briefly describing our approach for the preparation of two-qudit Bell-states (Bell$_d$). Noting that for qudit-dimension $d$, we define a \textit{Bell} state to be the pure and maximally entangled state given as
\begin{equation}
    \ket{\mathrm{Bell}_d} = \frac{1}{\sqrt{d}} \sum_{k=0}^{d-1} e^{i\phi_{k}} \ket{kk}
\end{equation}
where we allow $\phi_k \in [0,2\pi]$ to be an arbitrary phase on each state $\ket{kk}$ for $k = \{0,1,\dots,d-1\}$.

The Bell$_d$ states can be compactly generated from the vacuum (ground) state $\ket{00}$ with selective two-photon drives between $\ket{k,k} \leftrightarrow \ket{k+1,k+1}$. In particular, a Bell$_2$ state is realized via a single $\pi$ rotation of the $\ket{00}\leftrightarrow \ket{11}$ transition. A Bell$_3$ state is formed by a $2\pi/3$ rotation between $\ket{00}\leftrightarrow\ket{11}$ followed by a $\pi$ rotation between $\ket{11}\leftrightarrow\ket{22}$, where the state progresses (up to arbitrary phases $\phi_k$) as $\ket{00} \rightarrow  \sqrt{\frac{1}{3}}\ket{00} + \sqrt{\frac{2}{3}}\ket{11} \rightarrow \frac{1}{\sqrt{3}}(\ket{00} + \ket{11} + \ket{22}) $. Similarly, Bell$_4$ requires a $3\pi/4$ rotation between $\ket{00}\leftrightarrow\ket{11}$, followed by a $2\pi/3$ rotation between $\ket{11}\leftrightarrow \ket{22}$, and finally a $\pi$ rotation between $\ket{22}\leftrightarrow\ket{33}$. The experimental results for the tomographically reconstructed Bell$_d$ states for $d=\{2,3,4\}$ can be observed in Fig.~4 in the main text.

\begin{figure}[t]
    \centering
    \includegraphics{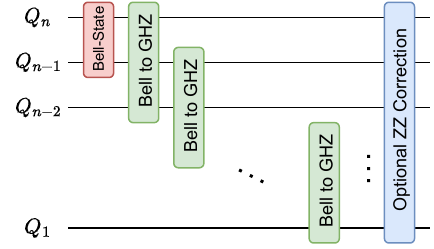}
    \caption{\textbf{Bell-to-GHZ generalization}. A qudit circuit of arbitrary size that prepares the GHZ state with the terminating correction or some similar maximally entangling state without it. After initializing a Bell state, this circuit repeatedly applies the Bell-to-GHZ kernel building up entangling one qudit at a time.}
    \label{fig:ghz-circuit} 
\end{figure}

\begin{figure*}[t]
    \centering
    \includegraphics[width=0.8\textwidth]{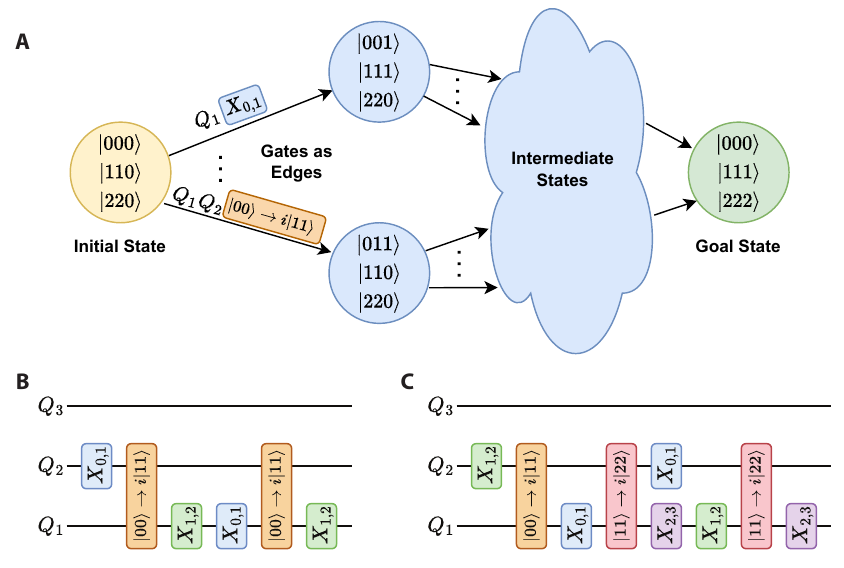}
    \caption{\textbf{Algorithm for circuit compilation}. (\textbf{A}) Algorithm workflow: the quantum state preparation problem can be simplified to a shortest path search when no parameterized gates are necessary. Each state becomes a node. Applying gates transitions one node to another, linking them in the graph. We attribute a cost to each gate and utilize Dijkstra's algorithm to search for the best path to a target state from some initial state. (\textbf{B}) Kernel for generating qutrit ($d=3$) GHZ states. (\textbf{C}) Kernel for generating ququart ($d=4$) GHZ states.}
    \label{fig:ghz_compilation}
\end{figure*}
\subsubsection{Kernel cycling: Modular approach to state preparation}

\noindent Next, we decompose the problem of GHZ state preparation into many applications of a three-qudit kernel. We designed this kernel to transition a Bell$_d$ state to a GHZ$_d$ state on three qudits, allowing us to cycle the application to build a GHZ state preparation circuit of any size, as shown in Fig.~S\ref{fig:ghz-circuit}. We start by preparing a qudit Bell$_d$ state on the first pair of qudits as outlined in the previous section. Then, on the first three qudits, we apply this transition defined by:

\begin{equation}
    \sum_{k<d}{\ket{k,k,0}} \rightarrow \sum_{k<d}{\ket{k,k,k}}.
\end{equation}

Afterward, we can cycle out the first qudit and cycle in the next one, resetting the system to the Bell$_d$ state. It is important to note that we are ignoring leading phases throughout this process. These can be corrected at the end of the circuit if desired.

\subsubsection{Kernel synthesis: Dijkstra's algorithm}

\noindent We desire a synthesized circuit that is composed without the use of Z rotations to minimize error accumulation. This restriction rules out the usage of state-of-the-art numerical synthesizers~\cite{younis2021berkeley}, as they require parameterized gates to be effective. 

Instead, we mapped our synthesis problem to a shortest path search by embedding the possible circuits as paths in a directed graph, as shown in Fig.~S\ref{fig:ghz_compilation}\textbf{A}. Here, nodes are unparameterized states, and edges are gates that transition one state to another. 

We utilize Dijkstra's algorithm to search for the shortest path to a GHZ state given the bell state starting node. We assign a cost to each edge such that the graph has no negative weights, enabling Dijkstra's algorithm to find the shortest path quickly. 
% We provide reference code as well as our results online at: \url{https://github.com/edyounis/Qudit-State-Transitions}.

\subsubsection{Preparation of GHZ states}

\noindent After applying the previous method to find solutions for qutrit, ququart, and ququint GHZ kernels, we discovered a pattern. In Fig.~S\ref{fig:ghz_compilation}\textbf{B,C}, we display the qutrit and ququart kernels. Here, we constructively prove that one can build a $d$-dimensional Bell to GHZ qudit state transition in $d-1$ two-qudit gates. We start with the qudit Bell state, $\sum_{k<d}{\ket{k,k,0}}$. In the first step, we simply apply an X gate to the $d-3$ and $d-2$ subspaces in the middle qudit, giving us:

\begin{equation}
\begin{split}
    \sum_{k<(d-3)}{\ket{k,k,0}} & + \ket{(d-3),(d-2),0} \\ & + \ket{(d-2),(d-3),0} \\ & + \ket{(d-1),(d-1),0}.
\end{split}
\end{equation}

Next, for all $l < d - 3$, we apply a two-photon transition, $\ket{l,l} \rightarrow i\ket{(l+1),(l+1)}$ to the last two qudits, followed by an X gate to the last qudit in the $l$ and $l+1$ subspaces,

\begin{equation}
\begin{split}
    \sum_{k<(d-2)}{\ket{k,(k+1),k}} & + \ket{(d-2),(d-3),(d-3)} \\ & + \ket{(d-1),(d-1),(d-3)}.
\end{split}
\end{equation}

Then, we apply another two-qudit transition $\ket{(d-3),(d-3)} \rightarrow i\ket{(d-2),(d-2)}$ on the last two qudits, followed by a sequence of X gates on the middle qudit. Counting $l$ up from zero to  $d-3$, we apply an X gate in the $l$ and $l+1$ subspaces:

\begin{equation}
\begin{split}
    \sum_{k<(d-3)}{\ket{k,k,k}} & + \ket{(d-3),(d-2),(d-3)} \\ & + \ket{(d-2),(d-2),(d-2)} \\ & + \ket{(d-1),(d-1),(d-3)}.
\end{split}
\end{equation}

Finally, we apply two X gates on the last qudit in subspaces $(d-2, d-1)$ and $(d-3, d-2)$, followed by a transition $\ket{(d-3),(d-3)} \rightarrow i\ket{(d-2),(d-2)}$. One more X gate on the rightmost qudit in the $(d-2, d-1)$ subspaces completes the algorithm to reach $\sum_{k<d}{\ket{k,k,k}}$.

This construction requires $d-1$ two-photon transitions. Furthermore, an $n$-qudit circuit will require the application of $n-2$ of these kernels alongside a bell-state preparation. The bell state preparation requires $d-1$ two-qudit gates, implying the total number of transitions for a GHZ circuit is $(n-1)(d-1)$, making it bilinear in qudit count and radix.

\end{document}